\documentclass[fleqn,usenatbib]{mnras}

\usepackage{newtxtext,newtxmath}

\usepackage[T1]{fontenc}

\DeclareRobustCommand{\VAN}[3]{#2}
\let\VANthebibliography\thebibliography
\def\thebibliography{\DeclareRobustCommand{\VAN}[3]{##3}\VANthebibliography}

\usepackage{graphicx}	
\usepackage{amsmath}	
\usepackage{amssymb}	

\usepackage{caption}
\usepackage{subcaption}
\usepackage{xcolor}

\def\xirppi{\xi^s(r_p,\pi)}
\def\xil{\xi_{\ell}}
\def\rpfc{r_p^{\left(\rm{fc}\right)}}
\def\({\left(}
\def\){\right)}
\def\[{\left[}
\def\]{\right]}

\def\mhmpc{\,h^{-1}{\rm Mpc}}

\def\mgpcc{\,{\rm Gpc^3}}
\def\mhgpcc{\,h^{-3}{\rm Gpc^3}}

\def\mwp{w_p(r_p)}

\newcommand{\aperp}{\alpha_\perp}	
\newcommand{\apara}{\alpha_\parallel}

\title[RSD from small-scale clustering of eBOSS LRGs]{The completed SDSS-IV extended Baryon Oscillation Spectroscopic Survey: measurement of the growth rate of structure from the small-scale clustering of the luminous red galaxy sample}

\author[M. J. Chapman et al.]{\parbox{\textwidth}{
Michael J. Chapman,$^{1,2}$\thanks{E-mail: mj3chapm@uwaterloo.ca}
Faizan G. Mohammad,$^{1,2}$
Zhongxu Zhai,$^{3}$
Will J. Percival,$^{1,2,4}$
Jeremy L. Tinker,$^{5}$
Julian E. Bautista,$^{6,7}$
Joel R. Brownstein,$^{8}$
Etienne Burtin,$^{9}$
Kyle S. Dawson,$^{8}$
Héctor Gil-Marín,$^{10,11}$
Axel de la Macorra,$^{12}$
Ashley J. Ross,$^{13}$
Graziano Rossi,$^{14}$
Donald P. Schneider,$^{15,16}$
Gong-Bo Zhao$^{17}$
 } \vspace*{4pt}\\
$^{1}$ Waterloo Centre for Astrophysics, University of Waterloo, Waterloo, ON N2L 3G1, Canada \\ 
$^{2}$ Department of Physics and Astronomy, University of Waterloo, Waterloo, ON N2L 3G1, Canada \\
$^{3}$ IPAC, California Institute of Technology, Mail Code 314-6, 1200 E. California Blvd., Pasadena, CA 91125 \\
$^{4}$ Perimeter Institute for Theoretical Physics, 31 Caroline St. North, Waterloo, ON N2L 2Y5, Canada \\
$^{5}$ Center for Cosmology and Particle Physics, Department of Physics, New York University, 726 Broadway, New York, NY 10003, USA\\
$^{6}$Institute of Cosmology and Gravitation, Dennis Sciama Building, University of Portsmouth, Portsmouth PO1 3FX, UK\\
$^{7}$Aix Marseille University, CNRS/IN2P3, CPPM, Marseille, France\\
$^{8}$ Department of Physics and Astronomy, University of Utah, 115 S. 1400 E., Salt Lake City, UT 84112, USA\\
$^{9}$ IRFU, CEA, Université Paris-Saclay, F-91191 Gif-sur- Yvette, France\\
$^{10}$ Institut de Ciències del Cosmos, Universitat de Barcelona, ICCUB, Martíı i Franquès 1, E08028 Barcelona, Spain\\
$^{11}$ Institut d’Estudis Espacials de Catalunya (IEEC), E08034 Barcelona, Spain\\
$^{12}$ Instituto de Física, Universidad Nacional Autónoma de México, Apdo. Postal 20-364, México\\
$^{13}$ Department of Astronomy, The Ohio State University, 140 W. 18th Ave., Columbus, OH 43210, USA\\
$^{14}$ Department of Physics and Astronomy, Sejong University, Seoul, 143-747, Korea\\
$^{15}$ Department of Astronomy and Astrophysics, The Pennsylvania State University, University Park, PA 16802\\
$^{16}$ Institute for Gravitation and the Cosmos, The Pennsylvania State University, University Park, PA 16802\\
$^{17}$ National Astronomical Observatories of China, Chinese Academy of Sciences, 20A Datun Road, Chaoyang District, Beijing 100012, China
}

\date{Accepted XXX. Received YYY; in original form ZZZ}

\pubyear{2022}

\begin{document}
\label{firstpage}
\pagerange{\pageref{firstpage}--\pageref{lastpage}}
\maketitle

\begin{abstract}
We measure the small-scale clustering of the Data Release 16 extended Baryon Oscillation Spectroscopic Survey Luminous Red Galaxy sample, corrected for fibre-collisions using Pairwise Inverse Probability weights, which give unbiased clustering measurements on all scales. We fit to the monopole and quadrupole moments and to the projected correlation function over the separation range $7-60\mhmpc$ with a model based on the \textsc{aemulus} cosmological emulator to measure the growth rate of cosmic structure, parameterized by $f\sigma_8$. We obtain a measurement of $f\sigma_8(z=0.737)=0.408\pm0.038$, which is $1.4\sigma$ lower than the value expected from 2018 Planck data for a flat $\Lambda$CDM model, and is more consistent with recent weak-lensing measurements. The level of precision achieved is 1.7 times better than more standard measurements made using only the large-scale modes of the same sample. We also fit to the data using the full range of scales $0.1-60\mhmpc$ modelled by the \textsc{aemulus} cosmological emulator and find a $4.5\sigma$ tension in the amplitude of the halo velocity field with the Planck+$\Lambda$CDM model, driven by a mismatch on the non-linear scales. This may not be cosmological in origin, and could be due to a breakdown in the Halo Occupation Distribution model used in the emulator. Finally, we perform a robust analysis of possible sources of systematics, including the effects of redshift uncertainty and incompleteness due to target selection that were not included in previous analyses fitting to clustering measurements on small scales.
\end{abstract}



\section{Introduction}

Understanding the accelerating expansion of the Universe is one of the primary goals for modern physics experiments. Many of these experiments aim to accomplish this through measuring the observed positions of galaxies in the Universe, which depend on the cosmological model in a number of ways. The intrinsic distribution of galaxies results from the growth of initial matter perturbations through gravity, giving a window to the early Universe. However, the fundamental observables are the angular positions and redshifts of galaxies, while the intrinsic pattern is in comoving distances, so surveys are also sensitive to the link between these two coordinates. This link depends on the relationship between separations in angles and redshifts and distances across and along the line of sight \citep{AP79}, as well as on redshift-space distortions \citep{Kaiser:1987}. Because these depend on both cosmological expansion and the build-up of structure within the Universe, large galaxy surveys offer a unique opportunity to solve the question of the origin of the late acceleration of the expansion \citep{weinberg_observational_2013, ferreira_cosmological_2019}.

The growth of structure most clearly manifests on the observed galaxy distribution through Redshift Space Distortions \citep[RSD]{Kaiser:1987}. These are a consequence of the velocities of galaxies in a comoving frame distorting the line-of-sight cosmological distances based on observed redshifts, and are sensitive to the growth rate of structure, which in turn depends on the strength of gravity. The strength of the RSD measurements depend on the parameter $f\sigma_8$, which is commonly used to quantify the amplitude of the velocity power spectrum and provides a strong test of modifications to gravity \citep{guzzo_test_2008,song_reconstructing_2009}. The development of large galaxy surveys driven by advances in multi-object spectrographs has resulted in recent renewed interest in RSD including measurements from the WiggleZ \citep{blake_wigglez_2011}, 6dFGS \citep{beutler_6df_2012}, SDSS-II \citep{samushia_interpreting_2012}, SDSS-MGS \citep{Howlett_clustering_2015}, FastSound \citep{okumura_subaru_2016}, and VIPERS \citep{pezzotta_vimos_2017} galaxy surveys. 

The best precision measurements to date come from the Baryon Oscillation Spectroscopic Survey (BOSS; \citealt{dawson_baryon_2013}), part of the third generation of the Sloan Digital Sky Survey (SDSS; \citealt{eisenstein_sdss-iii:_2011}). Using large-scale modes, BOSS has achieved the best precision of $\sim6$\% on the parameter combination $f\sigma_8$ \citep{beutler_clustering_2017, grieb_clustering_2017, sanchez_clustering_2017, satpathy_clustering_2017}. Note that these studies all measured RSD in the linear or quasi-linear regime, where proportionately small levels of non-linear modeling were required.

In contrast, \citet{Reid:2014} made a measurement of the amplitude of the RSD signal from an early BOSS galaxy sample, fitting to the monopole and quadrupole moments of the correlation function over scales $0.8$ to $32\mhmpc$, obtaining a 2.5\% measurement of $f\sigma_8(z=0.57)=0.450\pm0.011$. This demonstrates the increased precision available \emph{if} RSD in the data can be accurately measured and modeled to small scales. The most accurate method to model small-scale clustering is to use N-body simulations, and this was the route taken by \citet{Reid:2014}. However, without a simulation for each model to be tested (\citealt{Reid:2014} used three simulation sets at three very similar cosmologies), one has to extrapolate solutions to different cosmologies, which needs care. The most pernicious problem faced in the \citet{Reid:2014} analysis was correcting the small-scale clustering in the data, which suffers from fibre-collisions, where hardware limitations mean that some galaxies are excluded from the catalogue due to having close neighbours. A similar method was recently applied to the BOSS LOWZ galaxies \citep{Lange:2021}, and a study is in preparation for the CMASS sample \citep{Zhai_2021}.

The extended Baryon Oscillation Spectroscopic Survey (eBOSS; \citealt{dawson_sdss-iv_2016}), part of the SDSS-IV experiment \citep{blanton_sloan_2017} is the latest in a line of galaxy surveys made using the Sloan Telescope. This experiment was designed to make Baryon Acoustic Oscillations (BAO) and RSD measurements using three classes of galaxies used to directly trace the density field, together with a high redshift quasar sample \citep{duMasdesBourboux:2020} that allows Lyman-$\alpha$ forest measurements at redshifts $z>2.1$. We use the Luminous Red Galaxy (LRG) sample from Data Release 16 \citep{DR16} to make RSD measurements at $z\sim0.7$ including small-scale information. Standard BAO and RSD measurements made with this sample on larger scales only are presented in \citet{LRG_corr, gil-marin20a}, together with a test of their methodology using mock catalogues in \citet{Rossi:2020}. At intermediate redshifts, eBOSS probes the Universe using samples of emission line galaxies \citep{raichoor20a,tamone20a,demattia20a} and quasars \citep{Ross20a,lyke20a,hou20a,neveux20a,Smith:2020} as direct tracers of the density field lower redshifts. We do not analyse these data, focusing instead on the easier to model LRG sample. The cosmological interpretation of the BAO and RSD results from all eBOSS samples was presented in \citet{eBOSS_Cosmology}. 

Pushing the modelling to include small scales in our analysis is made possible by two key advances in methodology since the \citet{Reid:2014} analysis. First, we use the \textsc{aemulus} emulator \citep{zhai_aemulus3} to create accurate models of the redshift-space correlation function moments to small scales (see Section~\ref{sec:aemulus}). To correct for fibre-collisions, we use the Pairwise Inverse Probability (PIP) method \citep{Bianchi:2017,Percival:2017}, as described in Section~\ref{sec:pip}. Together, these advances mean that we can now both make and model accurate clustering measurements from the eBOSS LRG sample, fitting the correlation function to small scales.

Our paper is structured as follows: the eBOSS LRG sample is described in Section~\ref{sec:data}, and the method for measuring and fitting the correlation functions in Section~\ref{sec:method}. In Section~\ref{sec:model_tests} we perform various tests of the method using mock catalogues. We present our results in Section~\ref{sec:results}, and discuss their significance in Section~\ref{sec:discussion}. Finally, we summarize our results in Section~\ref{sec:summary}.

\section{eBOSS LRG sample} \label{sec:data}

The eBOSS LRG target sample was selected \citep{Prakash:2016} from SDSS DR13 photometry \citep{Albareti:2017}, together with infrared observations from the WISE satellite \citep{Lang:2016}. LRG targets were selected over 7500\,deg$^{2}$, and observed using the BOSS spectrographs \citep{Smee:2013} mounted on the 2.5-meter Sloan telescope \citep{Gunn:2006}. 

In order to measure clustering we quantify the sample mask, detailing where we could observe galaxies, using the random catalogue with 50 times more points than galaxies as described in \citet{Ross20a}. Regions with bad photometric properties, that are close to higher-priority targets, or near the centerpost region of the plates are masked, removing 17 per cent of the initial footprint. Redshifts for the randoms were sampled from those of the galaxies.

Redshifts were measured from the resulting spectra using the \textsc{redrock} algorithm\footnote{Available at github.com/desihub/redrock}. \textsc{redrock} fits the data with templates derived from principal component analysis of SDSS data, followed by a redshift refinement procedure that uses stellar population models. We are unable to obtain a reliable redshift estimate from many spectra ($3.5$ per cent on average across the survey), with a failure fraction with systematic angular variations. We therefore apply a weight $w_{\rm noz}$ as described in \citet{Ross20a} to galaxies to remove these variations, calculated as a function of position of the fibre on the detector and the signal-to-noise of that set of observations. 

\begin{figure}
	\includegraphics[width=\columnwidth]{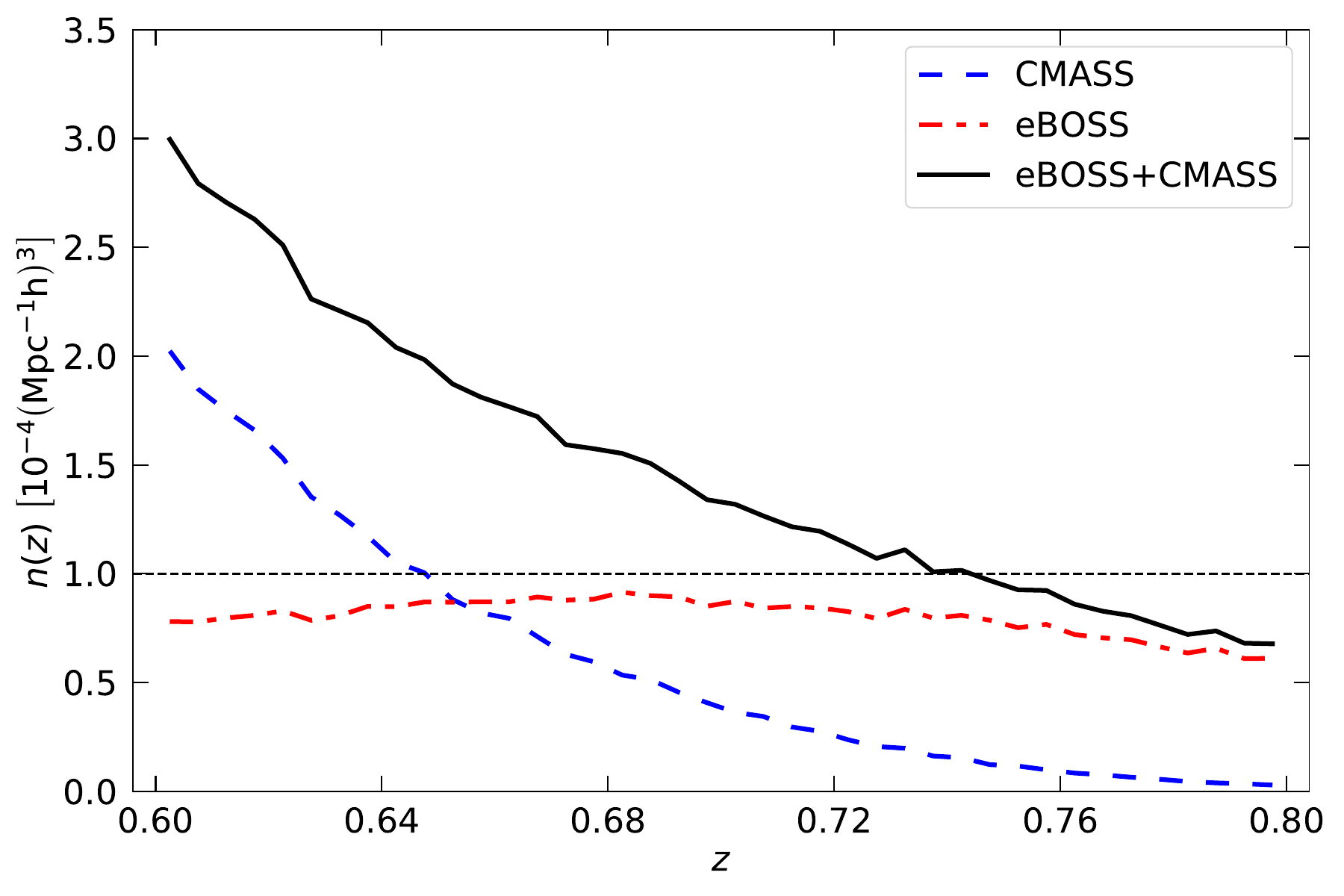}
    \caption{Redshift distribution of the eBOSS DR16 (red dash-dotted line), CMASS DR12 (blue dashed line) and the joint eBOSS+CMASS sample (black thick line, see Sec.~\ref{sec:cmass_eboss_results} for details), optimized using $w_{\rm FKP}$ weights.}
    \label{fig:cmass_eboss_nz}
\end{figure}

Systematic variations in the density of galaxies caused by variations in the photometric data used for target selection are mitigated by applying weights $w_{\rm sys}$ to the galaxies. These were computed using a multi-linear regression on the observed relations between the angular over-densities of galaxies versus stellar density, seeing and galactic extinction. As we are interested primarily in small-scales, the exact correction is not important. Additional weights $w_{\rm FKP}$ that optimise the signal, which varies because the density varies across the sample \citep{Feldman:1994}, are also included (Fig.~\ref{fig:cmass_eboss_nz}).

A fibre could not be placed on 4 per cent of the LRG targets due to fibre-collisions: when a group of two or more galaxies are closer than 62$^{\prime\prime}$ they cannot all receive a fibre because of hardware limitations. We use Pairwise Inverse Probability (PIP) weights $w^\mathrm{PIP}$ together with angular upweighting \citep{Bianchi:2017,Percival:2017} to correct for this effect, as described in \citet{Mohammad:2020}, and section~\ref{sec:pip}. The final combined weight applied to the galaxies is defined as $w^\mathrm{tot} = w^{\rm noz}w^{\rm sys} w^{\rm FKP}$, and we also use $w^\mathrm{PIP}$ applied to pairs.

The eBOSS sample of LRGs overlaps in area and redshift range with the high-redshift tail of the BOSS CMASS sample. Unlike many other eBOSS analyses including the large-scale measurements of BAO and RSD \citep{LRG_corr,gil-marin20a}, we do not combine the eBOSS LRG sample with all the $z > 0.6$ BOSS CMASS galaxies. We focus on the eBOSS sample to simplify the correction of the small-scale fibre assignment: fibre assignment was performed separately for BOSS and eBOSS using different configurations of the SDSS tiling code.

We define the effective redshift of our sample as the weighted mean redshift of galaxy pairs,
\begin{equation}
  z_{\mathrm{eff}}=\frac{\Sigma_{m>n}w^\mathrm{PIP}_{mn}w^\mathrm{tot}_n w^\mathrm{tot}_n(z_m+z_n)/2}
          {\Sigma_{m>n}w^\mathrm{PIP}_{mn}w^\mathrm{tot}_mw^\mathrm{tot}_n},
\end{equation}
where the indices $m$, $n$ are over the objects in the data catalogue, and the description of the weights is given in Sec.~\ref{sec:pip}. Additionally, we only include galaxy pairs which have a separation between $0.1-60\mhmpc$, the scales used in our measurement. The effective redshift we obtain for our sample is $z=0.737$, and an effective comoving volume of $1.28\mgpcc$ \citep{Ross20a}.

\section{Methods}  \label{sec:method}

\subsection{Measurements}

We measure and model the observed galaxy clustering in redshift space using the two-point correlation function as calculated using the least-bias and least-variance Landy-Szalay estimator \citep{landy93},
    \begin{equation}
        \xi\left(\boldsymbol{s}\right) = \frac{DD\left(\boldsymbol{s}\right)-2DR\left(\boldsymbol{s}\right)}{RR\left(\boldsymbol{s}\right)}+1, \label{eq:LS}
    \end{equation}
with $DD$, $DR$ and $RR$ being the data-data, data-random and random-random pair counts at a given separation $\boldsymbol{s}$. To reduce the impact of shot noise on the measured $\xi$ from the random catalogue we use a number of random points $N=50$ times the number of galaxies in the DR16 sample. The difference in the number of galaxies and randoms is accounted for by normalising the pair counts in Eq.~\ref{eq:LS} to the total number of distinct pairs.

The modelling of the 3D correlation function in Eq.~\ref{eq:LS} is complicated by the large number of separation bins. Indeed, this requires a very large number of survey realizations to estimate the data covariance matrix. We follow the standard technique of compressing the information contained in the full 3D correlation function $\xi\left(\boldsymbol{s}\right)$. In particular we fit our model to the projected correlation function $\mwp$ and the first two even multipole moments $\xil$ of the redshift space correlation function.

The halo-occupation properties of a given sample affect its intrinsic clustering. Classically, this effect is modelled using the projected correlation function $\mwp$ that is expected to be free of the apparent RSD effects. The projected correlation function $\mwp$ is estimated through,
    \begin{equation}
        w_p\left(r_p\right) = 2\int_{0}^{\pi_{\mathrm{max}}}\xirppi d\pi, \label{eq:wp}
    \end{equation}
where $r_p$ and $\pi$ are the normal and parallel to the line-of-sight (los) components of the pair separation $\boldsymbol{s}$. We limit the integral in Eq.~\ref{eq:wp} to a maximum los separation of $\pi_{\mathrm{max}}=80\mhmpc$, matching the definiton in the model to be fitted to these data \citep{zhai_aemulus3}.

Redshift-space distortions change the apparent positions of targets in the radial direction with respect to those in real-space. RSD are classically measured and modelled in the multipole moments $\xil$ of the redshift-space correlation function $\xirppi$. Multipole moments $\xil$ are defined as,
    \begin{equation}
        \xil(s) = \left( 2\ell+1 \right)\int_{0}^{1}\xi^{s}\left(s,\mu\right)L_\ell(\mu)d\mu, \label{eq:mps}
    \end{equation}
with $s=|\boldsymbol{s}|$ and $\mu=\pi/s$ is the cosine of the angle between the los direction and the pair separation vector $\boldsymbol{s}$ and $L_\ell$ is the $\ell$-order Legendre polynomial.

We bin $r_p$ and $s$ in 9 logarithmically spaced bins between $0.1-60\mhmpc$,  matching the output of \textsc{aemulus} predictions for $\mwp$ and $\xil$, while the los separation $\pi$ and $\mu$ are binned using linear bins of width $\Delta\pi=1\mhmpc$ and $\Delta\mu=0.1$. Given the discrete binning of different variables we estimate the integrals in Eq.~\ref{eq:wp} and Eq.~\ref{eq:mps} as Riemann sums.

\subsection{PIP Correction}  \label{sec:pip}

In eBOSS spectroscopic observations, fibre-collisions occur whenever two targets are closer than $\theta^{(\rm{fc})}=62\arcsec$ on the sky. While a fraction of such collisions are resolved thanks to multiple passes of the instruments in small chunks of the survey, fibre-collisions in single passes remain un-resolved and correlate with the underlying target density. If not properly corrected, missed targets due to fibre-collisions can systematically bias the measured two-point correlation function on small scales. In the large scale analysis of the eBOSS LRG sample \citep{LRG_corr} fibre-collisions are accounted for by means of the nearest-neighbour (NN) weighting that is quantified through the weight $w_{\rm{cp}}$.

In this work we replace the standard NN correction for fibre-collisions with a more rigorous Pairwise-Inverse-Probability (PIP) weighting \citep[see][for a discussion about inverse-probability estimators]{Bianchi:2017}. The PIP weights are assigned to pairs of objects in the targeted sample and quantify the probability, for any pair, of being targeted in a random realisation of the survey targeting. Under the assumption that no pair has zero probability of being observed, applying the PIP weighting provides statistically un-biased estimates of the two-point correlation function. The selection probabilities are characteristic of the particular fibre assignment algorithm used to select targets from a parent photometric sample for the spectroscopic follow-up. Therefore, these probabilities are extremely difficult to model analytically except for some simple targeting strategies. We infer the selection probabilities by generating multiple replicas of the survey target selection. Details on how these survey realisations are built are provided in \citet{Mohammad:2020}. Given a set of survey realisations, the inverse probability, or equivalently the PIP weight $w_{mn}$, is simply the number of realisations in which a given pair could have been targeted divided by the number of times it was targeted. The individual-inverse-probability (IIP) $w_m$ are the single-object counterparts of the PIP weights, i.e. the inverse-probability for a given object $m$ of being targeted in a random survey realisation.

PIP weighting assumes that all pairs have a non-zero chance of being observed. However, in eBOSS pairs with separation smaller than the fibre-collision scale $\theta^{(\rm{fc})}$) are missed in single-pass areas in all survey realizations. These pairs produce a systematic underestimation in the measured two-point correlation function. For the eBOSS LRG sample, the systematic bias is confined at transverse scales smaller than $\rpfc\sim 0.7\mhmpc$ in $\mwp$ while it spreads to larger separations $s$ in the multipole moments $\xil$. Truncated multipoles $\hat\xi^{\left(\ell\right)}$ were proposed in \citet{Reid:2014, mohammad16} to remove transverse scales $r_p <\rpfc$ from the measured multipole moments, resulting in a loss of information at scales smaller than $\rpfc$. Alternatively, the angular up-weighting outlined in \citet{Percival:2017} can be used to de-bias the measurements at smaller scales. The angular up-weighting relies on the assumption that pairs missed due to fibre-collisions in the single pass of the instrument are statistically equivalent to those targeted in the multiple-pass areas. The combined PIP and angular up-weighting (PIP+ANG) is,
\begin{equation}{\label{eq:pip+ang}}
\left.\begin{aligned}
    DD(\vec{s}) &= \sum_{\substack{\vec{x}_m - \vec{x}_n \approx \vec{s}\\\vec{u}_m\cdot \vec{u}_n\approx\cos{\theta}}} \mathrm{w}^{\mathrm{PIP}}_{mn}w^{\rm{tot}}_{m}w^{\rm{tot}}_{n}\times \frac{DD_{\rm{par}}\left(\theta\right)}{DD_{\rm{fib}}^{\rm{PIP}}\left(\theta\right)} \ ,\\
    DR(\vec{s}) &= \sum_{\substack{\vec{x}_m - \vec{y}_n \approx \vec{s}\\\vec{u}_m\cdot \vec{v}_n\approx\cos{\theta}}} \mathrm{w}^{\mathrm{IIP}}_{m}w^{\rm{tot}}_{m}w^{\rm{tot}}_{n}\times \frac{DR_{\rm{par}}\left(\theta\right)}{DR_{\rm{fib}}^{\rm{IIP}}\left(\theta\right)} \ ,
    \end{aligned}\right.
\end{equation}
where $w^{\rm{tot}}=w^{\rm{sys}} w^{\rm{noz}} w^{\rm{FKP}}$, and $w^{\mathrm{PIP}}_{mn}$ and $w^{\mathrm{IIP}}_m$ are PIP and IIP weights, respectively. The fractions on the right-hand side in Eq.~\ref{eq:pip+ang} are the angular weights for $DD$ and $DR$ pair counts. An extensive description of different terms in Eq.~\ref{eq:pip+ang} is given in \citet{Mohammad:2020}.

\citet{Mohammad:2020} extensively tested the effectiveness of the method of PIP+ANG weighting using a sample of 100 Effective Zel’dovich mocks \citep[EZmocks,][]{zhao20} designed to match the eBOSS LRG sample. The mean of the corrected measurements was compared to the mean of the true clustering of the mocks for $\xi_0$, $\xi_2$ and $w_p$ over a separation range of $0.1-100\mhmpc$ \citep[see Fig. 9 and 12 of ][]{Mohammad:2020}. The PIP+ANG correction was able to recover the clustering of the parent sample to within 1-$\sigma$ of the error on the mean at all measurement scales for $\xi_0$, and $\xi_2$ and all scales of $w_p$ except for the fibre-collision scale, where the corrected measurements recovered the true clustering to within the error on a single mock. We can therefore be confident that the PIP+ANG correction to the eBOSS LRG sample produces unbiased results to within the statistical uncertainty of our sample on all scales.

\subsection{\textsc{aemulus} Cosmological Emulator}  \label{sec:aemulus}

We compare our measurements to the \textsc{aemulus} cosmological emulator \citep{zhai_aemulus3} predictions for $\xi_0$, $\xi_2$, and $w_p$ for a galaxy sample in a universe with variable cosmological and galaxy-halo connection parameters. The \textsc{aemulus} emulator applies Gaussian process based machine learning to a training set of 40 N-body simulations and that use a latin hypercube to optimally sample a wCDM parameter space spanning the approximate $4\sigma$ range of the Planck \citep{Planck-2018-params} or WMAP \citep{WMAP-final-params} results \citep{derose_aemulus1}. A halo occupation distribution model (HOD) is used to connect a galaxy sample to the dark matter halos. Unlike some galaxy clustering analyses, our emulator does not model $\xi_4$, since it is considerably noisier than $\xi_0$ and $\xi_2$. The emulator prediction would likely be noise dominated for $\xi_4$, and would require adding more training complexity without a commensurate increase in cosmological information. In their measurement of $f\sigma_8$ from small-scale clustering within the BOSS LOWZ sample, \citet{Lange:2021} found that excluding $\xi_4$ from their analysis of $\xi_0$ and $\xi_2$ did not produce a significant change in best fit value or uncertainty.

\textsc{aemulus} allows for a flat wCDM with described by 7 parameters: $\Omega_M$, $\Omega_b$, $\sigma_8$, $h$, $n_s$, $w$, and $N_{\mathrm{eff}}$. For our analysis we limit the cosmological parameter space by fixing $N_{\rm eff}=3.046$ and $w=-1$, since these parameters are not well constrained by our measurements but have been well measured by other probes, resulting in a 5 parameter flat $\Lambda$CDM cosmology. The HOD model used by the \textsc{aemulus} allocates a Poisson sampling of $N(M)$ galaxies to halos of mass $M$, split into central galaxies and satellites following:

\begin{equation}  \label{eq:hod_cen_sat}
    \langle N(M)\rangle  = N_{\rm cen}(M) + N_{\rm sat}(M),
\end{equation}

\begin{equation}  \label{eq:hod_cen}
    N_{\rm cen}(M) = \frac{f_{\rm max}}{2}\left[1 + {\rm erf}\left(\frac{\log_{10}M-\log_{10}M_{\rm min}}{\sigma_{\log M}} \right)\right],
\end{equation}

\begin{equation}  \label{eq:hod_sat}
    N_{\rm sat}(M) = \left(\frac{M}{M_{\rm sat}}\right)^{\alpha}\exp \left(-\frac{M_{\rm cut}}{M} \right)\frac{N_{\rm cen}(M)}{f_{\rm max}}.
\end{equation}

where the free parameters fit by the emulator are $f_{\rm max}$, $\sigma_{\log M}$, $\log M_{\mathrm{sat}}$, $\alpha$, $\log M_{\mathrm{cut}}$. Briefly, $\sigma_{\log M}$ defines the width of the transition from a mean occupation of 0 to 1 for centrals, $M_{\mathrm{sat}}$ is the typical mass for halos to host one satellite, $\alpha$ is the power-law index for the mass dependence of the satellite occupation, and $M_{\mathrm{cut}}$ gives an exponential cutoff to the satellite occupation at low mass. $M_{\mathrm{min}}$ sets the transition point of the central occupation, and is fixed in the emulator to match the number density of the sample. By matching the number density we ensure the correct linear bias, thus reducing the degeneracy between the HOD parameters and the growth rate in the correlation function measurements. Because of this choice we do not use the number density as an observable in our analysis. $f_{\rm max}$ is a new parameter that we add to \textsc{Aemulus} to address a possible inconsistency between the model and data. eBOSS was targeted using colour and magnitude cuts \citep{Prakash:2016} so it is not a complete sample, whereas the HOD model assumes that all galaxies are included in the sample. This is especially concerning for eBOSS since targets were selected using a lower magnitude limit in the i band to avoid overlap with the CMASS LRG sample (see Fig.1 of \citealt{Zhai:2017}). $f_{\rm max}$ controls the fraction of centrals that are included in the sample, i.e. a value of $f_{\rm max}<1$ means that the very massive halos do not necessarily host a eBOSS LRG at the center. While these targeting cuts would be expected to affect the completeness of both centrals and satellites, for satellites $f_{\rm max}$ is completely degenerate with $M_{\mathrm{sat}}^{-\alpha}$ \citep[see similar discussion in][]{Lange:2021}. Since these HOD parameters are primarily nuisance parameters in our constraint of the growth rate, we do not apply $f_{\rm max}$ to the satellites. In Sec.~\ref{sec:fmax_test} we perform a series of tests to determine the effect of excluding $f_{\rm max}$ on the measured $f\sigma_8$.

The emulator also allows three additional parameters that control how galaxies are distributed in their host halos: $c_{\rm vir}$, $v_{\rm bc}$, and $v_{\rm bs}$ (labelled $\eta_{\mathrm{con}}$, $\eta_{\mathrm{vc}}$, and $\eta_{\mathrm{vs}}$ in \citealt{zhai_aemulus3}). $c_{\rm vir}$ is the ratio between the concentration parameters of the satellites to the host halo where the halo is assumed to have a Navarro-Frenk-White (NFW) profile \citep{NFW}. $v_{\rm bc}$ and $v_{\rm bs}$ are the velocity biases of centrals and satellites respectively, where $\sigma_{\rm gal}=v_{\rm gal}\sigma_{\rm halo}$ and $\sigma_{\rm halo}$ is the velocity dispersion of the halo calculated from its mass. Finally, the \textsc{aemulus} emulator uses a 15th parameter, $\gamma_f$, which rescales all halo bulk velocities in the simulation. The galaxy velocity can therefore be thought of as the sum of two components: a component equal to the bulk motion of the host halo scaled by $\gamma_f$, and a randomly directed component that depends on the halo mass through the velocity dispersion and that is scaled by either $v_{\rm bc}$ or $v_{\rm bs}$ for centrals and satellites respectively. For a detailed description of the \textsc{aemulus} correlation function parameters see \citet{zhai_aemulus3}. See Sec.~\ref{sec:fit} for a description of how we treat these parameters in our fit.

The original \textsc{aemulus} emulator was trained to match a BOSS CMASS-like sample at $z=0.57$ and space density $n=4.2\times 10^{-4} [h^{-1}\text{Mpc}]^{-3}$. However, our eBOSS sample is at an effective redshift of $z=0.737$ and peak number density of $n=9\times 10^{-5}$. The difference in number density is particularly worrying, since a less dense sample will preferentially fill more massive halos. The result will be a sample with a larger linear bias, which is degenerate with the growth rate in clustering measurements. In order to ensure an unbiased result we rebuild the emulator from the original simulations, but using the $z=0.7$ simulation time-slice and adjusting HOD parameters, especially $M_{\mathrm{min}}$, to match the eBOSS number density. The training ranges for the new emulator are given in Table. ~\ref{table:parameters}.

\subsection{Interpreting growth rate measurements} \label{sec:growth_meas}

As shown in \citet{Reid:2014}, which used a similar parameterization to measure RSD from their simulations, in the linear regime a fractional change in $\gamma_f$ is proportional to a fractional change in $f$, such that $f=\gamma_f f_{\Lambda\mathrm{CDM}}$, where $f_{\Lambda\mathrm{CDM}}$ is the linear growth rate for a flat $\Lambda$CDM cosmology specified by the model parameters. However, the link between the linear velocity power spectrum amplitude and the non-linear regime is possibly scale dependent. I.e. a linear response on large scales might not necessarily lead to a linear response on small scales. $\gamma_f$ is introduced in the simulations as a scaling of all velocities by the same amount and so $\gamma_f$ also scales the non-linear velocities of halos. In this case $\gamma_f$ still provides a consistency test with the amplitude of the velocity field expected in a $\Lambda$CDM universe with the model cosmology, where $\gamma_f=1$ indicates agreement, but it no longer necessarily gives a pure rescaling of the linear growth rate. For models that do have such a linear response, then the measurement of $\gamma_f$ over the full range of scales can be used to constrain the linear growth rate. However, as this is model dependent, we conservatively separate the contributions of the linear and non-linear regime in presenting our results (as described in Sec.~\ref{sec:non-linear}).

Although the \textsc{aemulus} code uses $\gamma_f$ to adjust the RSD amplitude in the model, the RSD are sensitive to the parameter combination $f\sigma_8$. We therefore present our large-scale results in terms of $f\sigma_8=\gamma_f f_{\Lambda\mathrm{CDM}}\sigma_8$, which is used in the remainder of the paper and the abstract. It is also important to note that we calculate $f_{\Lambda\mathrm{CDM}}\sigma_8$ from the model cosmology according to linear theory, rather than the value that would be obtained from the power spectrum on scales corresponding to $0.1-60 \mhmpc$. Thus the value of $f\sigma_8$ we present is the value expected from linear theory for our model, and is directly comparable to measurements made on larger scales. However, care should be taken when using the resulting measurements of $f\sigma_8$ to constrain models where the other parameters deviate significantly from flat $\Lambda$CDM and general relativity ($\Lambda$CDM+GR, hereafter used interchangeably with $\Lambda$CDM). A problem inherent in many cosmological measurements and all previous RSD measurements is that one assumes various features of a particular model, here flat $\Lambda$CDM, in order to make the measurements. To test a different model, one should strictly have to perform a new fit including all properties of that model. This does not affect the validity of our measurement as a test of consistency with $\Lambda$CDM within the parameter space of the emulator, or as an indication of how the RSD measurements compare to those from other surveys.

\subsection{Covariance Matrix} \label{sec:cov-mat}

\begin{figure*}
	\includegraphics[width=\textwidth]{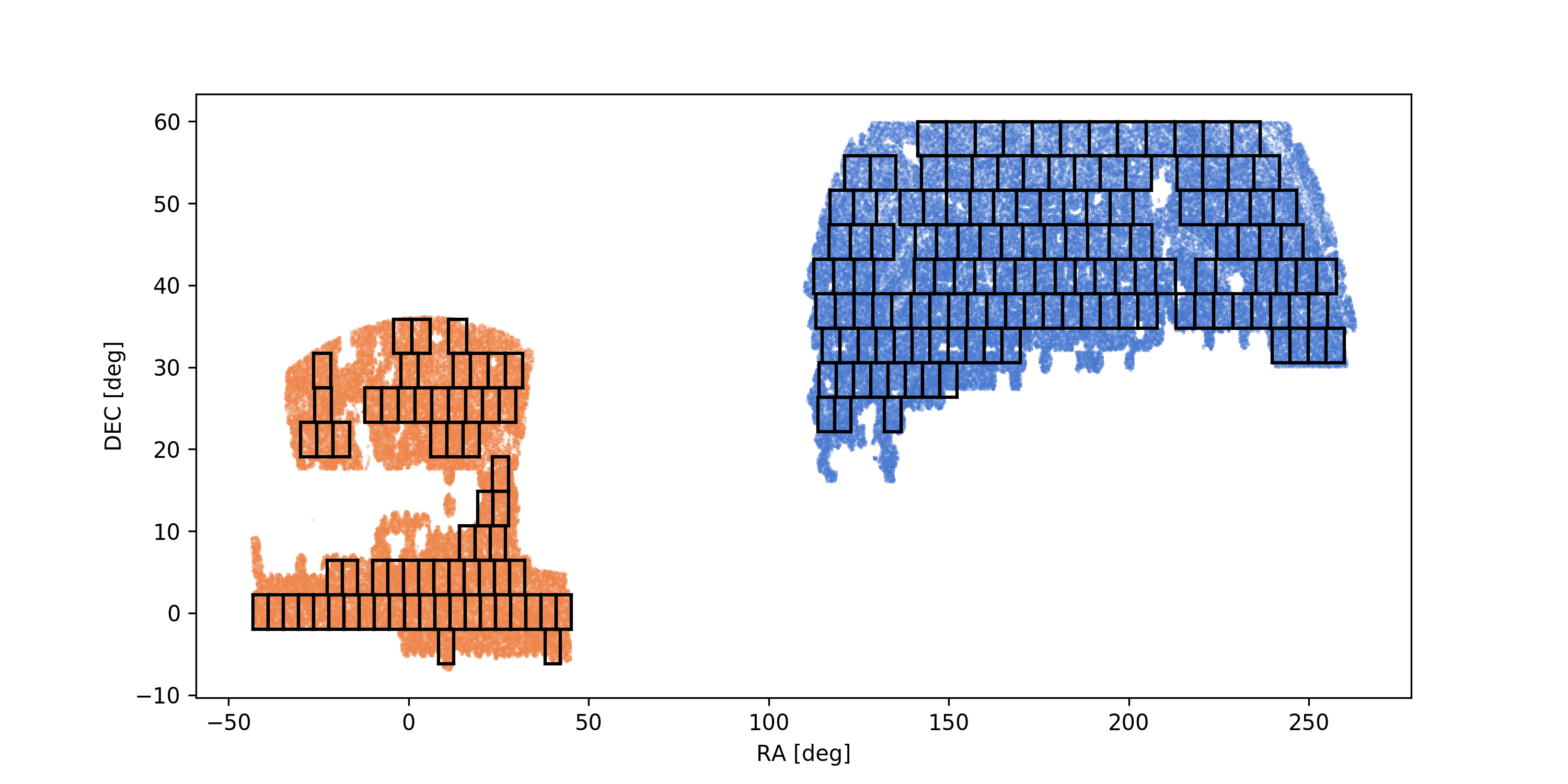}
    \caption{The footprint of the eBOSS LRG clustering catalogue with our jackknife regions. The blue points show the North Galactic Cap (NGC) observations, while the orange points show the South Galactic Cap (SGC) observations. It should be noted that the square jackknife regions all have approximately equal area on the sky, however due to the distortion of projecting a sphere onto a plane the regions at larger declination appear wider in this plot.}
    \label{fig:jk_division}
\end{figure*}

Clustering measurements in different separation bins are correlated, and we need an estimate of the covariance matrix when fitting a model to the observations. Mock surveys, either based on the output of N-body simulations or approximate methods, have been widely used to estimate the data covariance matrix. However, in order to work on small scales, we would need a large number of simulations that accurately reproduce the small-scale clustering - a difficult task. In order to generate a covariance matrix that reflects the small-scale clustering of our sample we instead use jackknife sampling. We split our survey footprint into equal area squares on the sky using right ascension (RA) and declination (DEC) cuts. This method relies on the clustering of the sample being uncorrelated with position in the survey. Furthermore, because we expect the covariance to follow a simple volume scaling, we remove the squares with the smallest occupation as determined from the random catalogue over the survey footprint, so that each region included will contribute approximately the same statistical weight to the sampling (Fig.~\ref{fig:jk_division}). Since the measurements from each sample are normalized it is not necessary that they contain identical numbers of objects, however selecting regions in this way reduces variance from regions at the edge of the survey which are only partially filled or have peculiar geometries. The missing area is included in the final calculation by means of a volume-weighted correction.

For the objects in our data and random catalogues that are located within one of the 200 accepted regions we store a region identification number. We then recalculate the monopole and quadrupole of the 3D correlation function and projected correlation function for this reduced sample 200 times, excluding one region from the calculation each time. We include the full PIP+ANG weighting scheme in these calculations, so that the variance in the PIP+ANG weights is included in the jackknife estimation. The covariance matrix is then estimated from this jackknife sampling using
\begin{equation}  \label{eq:jackknife}
    C_{i,j} = \frac{n-1}{n}\sum_{k}^{n}(\xi_{i,k} - \bar{\xi}_i)(\xi_{j,k} - \bar{\xi}_j),
\end{equation}
 where the $i,j$ indices are over the elements of the data vector, n=200 is the number of jackknife regions, and $k$ is an index over the jackknife realisations.

In order to more easily visualize the correlations between bins we calculate the correlation matrix by:
\begin{equation}  \label{eq:corr_mat}
    R_{i,j} = \frac{C_{i,j}}{(C_{i,i}C_{j,j})^{1/2}},
\end{equation}

\begin{figure}
	\includegraphics[width=\columnwidth]{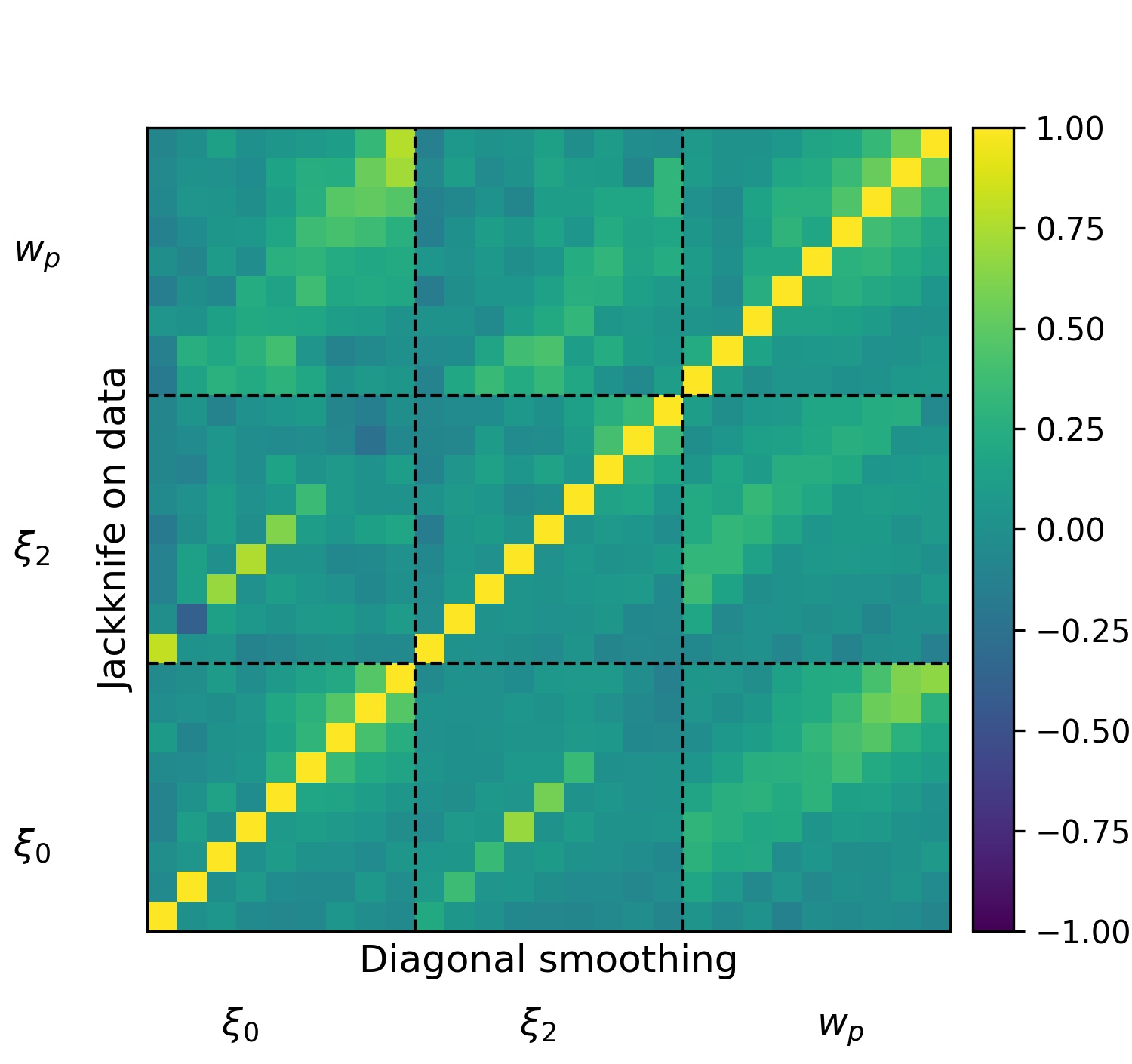}
    \caption{Comparison of the unsmoothed and smoothed correlation matrices. The upper diagonal elements correspond to the unsmoothed jackknife correlation matrix, while the lower diagonal elements show the result of our diagonal smoothing method.}
    \label{fig:cor_mat_smoothing}
\end{figure}

The correlation matrix is highly diagonal, which is expected since we have a small number of widely separated bins, which are only expected to be weakly correlated. In order to reduce the noise in the off-diagonal terms we smooth the correlation matrix using diagonally adjacent bins. Each off-diagonal element is assigned the average of itself and the two adjacent diagonal elements, excluding bins from other measurements. The result of this diagonal smoothing is shown in Fig.~\ref{fig:cor_mat_smoothing}. 

\begin{figure*}
	\includegraphics[width=\textwidth]{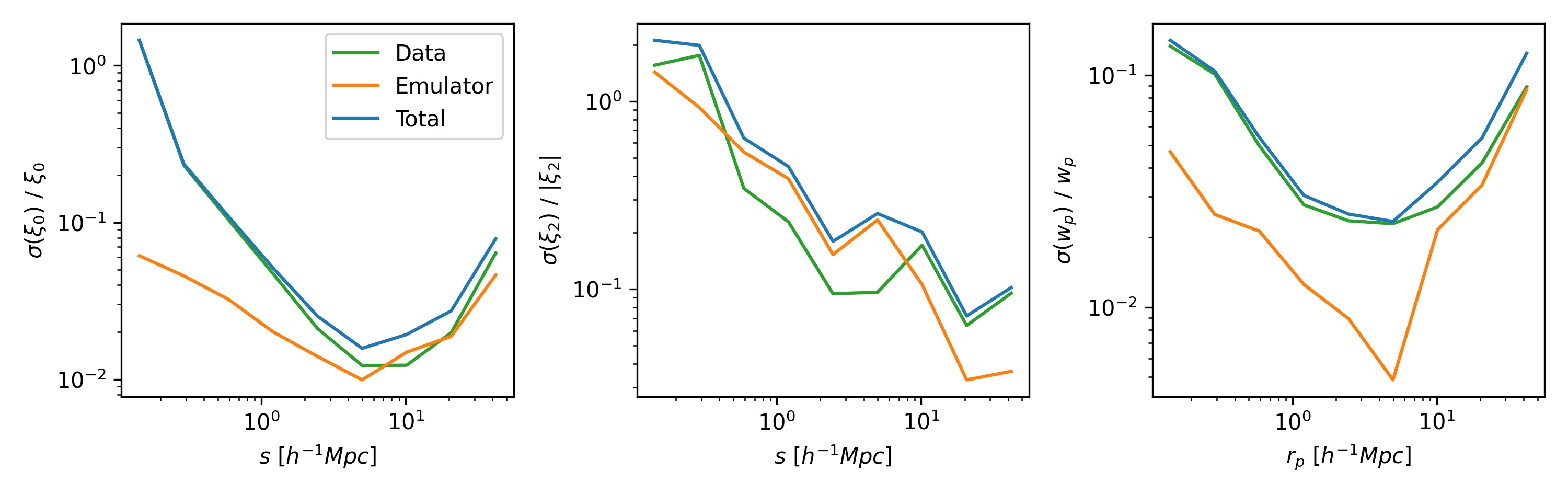}
    \caption{The contributions of the data error calculated through jackknife sampling (green), the emulator error (orange), and total error (blue), for the monopole, quadrupole, and projected correlation function (left to right).}
    \label{fig:error_components}
\end{figure*}

In addition to the data error we include the emulator error in the covariance matrix. The emulator error is calculated as a fractional error on each correlation function bin using a sample of test HOD parameter sets which are selected from the same parameter ranges as the training sample, but were not used in the training \citep{zhai_aemulus3}. The fractional error is converted to an absolute error, $\sigma_E$, by multiplying by the correlation function measurements from the data. The total variance for each measurement bin is then calculated from $\sigma_T^2=\sigma_D^2+\sigma_E^2$. In order to preserve the structure of the jackknife covariance matrix we convert the smoothed correlation matrix back to the covariance matrix using $C_{i,i}=\sigma_{T,i}^2$. The contributions of the data and emulator errors to the total error are shown in Fig.~\ref{fig:error_components}. The data error is dominant in the region $s<5\mhmpc$ for the monopole and projected correlation function, while the emulator error is comparable for $s>5\mhmpc$ and across the full separation range of the quadrupole.

We also correct the inverse covariance matrix according to \citet{Hartlap:2007}, using 
\begin{equation}  \label{eq:hartlap}
   \hat{\boldsymbol{C}}^{-1} = \frac{n-p-2}{n-1}\boldsymbol{C}^{-1},
\end{equation}
where $n=200$ is the number of jackknife regions, and $p=27$ is the number of combined bins in our three measurements. Although $n$ should properly be the number of completely independent measurements \citep{Krauss:2013, Eifler:2008}, we follow \citet{Reid:2014} in using the number of regions, noting that this correction may therefore underestimate the true size of the effect. However, this factor has very little effect on our final fit, as well as not changing the best fit value.

\subsection{AP Scaling}

Although we fit the \textsc{aemulus} correlation function predictions directly to our measurements from the data, our results are still affected by the Alcock-Paczynski (AP) effect \citep{AP79}, because we convert the data redshift to distance assuming a fixed fiducial cosmological model. We therefore need to scale the separations between model and data to account for the difference in comoving distance between our fiducial cosmology and the cosmology of the model. We apply the standard AP scaling from \citet{LRG_corr} to each model, first defining the perpendicular and parallel dilation factors
\begin{equation}
  \aperp = \frac{D_M(z_{\rm eff})}{D_M^{\rm fid}(z_{\rm eff})}\,,\hspace{0.5cm}
  \apara = \frac{D_H(z_{\rm eff})}{D_H^{\rm fid}(z_{\rm eff})}\,,
\end{equation}
where $D_M$ is the comoving angular diameter distance, $D_H$ is Hubble distance. We then scale the multipole moments of the correlation function as follows
\begin{equation}
    \xi_0^{\mbox{fid}}(r^{\mbox{fid}})=\xi_0(\alpha r) + \frac{2}{5}\epsilon\left[3\xi_2(\alpha r) + \frac{d\xi_2(\alpha r)}{d\ln(r)}\right]\,,
	\label{eq:standard_scaling_xi0}
\end{equation}
\begin{equation}
    \xi_2^{\mbox{fid}}(r^{\mbox{fid}})=(1+\frac{6}{7}\epsilon)\xi_2(\alpha r) + 2\epsilon\frac{d\xi_0(\alpha r)}{d\ln(r)} + \frac{4}{7}\epsilon\frac{d\xi_2(\alpha r)}{d\ln(r)}.
	\label{eq:standard_scaling_xi2}
\end{equation}
where $\alpha=\apara^{1/3}\aperp^{2/3}$ and $\epsilon=(\apara/\aperp)^{1/3}-1$. Once we have shifted the model, we used a cubic spline interpolation to recover the model values at the fiducial separations used to calculate the data values. 

The projected correlation function was calculated similarly using the scaling
\begin{equation}
    w_p^{\mbox{fid}}(r_p^{\mbox{fid}}) =  w_p(\aperp r_p)\,.
\end{equation}

The accuracy of this method depends in part on the width of the bins used due to the calculation of the derivative and the interpolation between points. In order to assess the importance of these factors we perform an additional fit to the data without the AP correction. See Sec.~\ref{sec:ap_results} for details.

\subsection{Exploring the Likelihood} \label{sec:fit}

\begin{table}
\centering
\begin{tabular}{|l|c|c|}
    \hline
    Parameter & Training Range & Prior Range\\
    \hline
    $\Omega_m$ & [0.255, 0.353] & [0.225, 0.375]\\
    $\Omega_b h^{2}$ & [0.039, 0.062] & [0.005, 0.1]\\
    $\sigma_8$ & [0.575, 0.964] & [0.5, 1]\\
    $h$ & [0.612, 0.748] & [0.58, 0.78]\\
    $n_s$ & [0.928, 0.997] & [0.8, 1.2]\\
    $N_{\mathrm{eff}}$ & [2.62, 4.28] & 3.046\\
    $w$ & [-1.40, -0.57] & -1\\
    \hline
    $\log M_{\mathrm{sat}}$ & [14.0, 16.0] & [13.8, 16.2]\\
    $\alpha$ & [0.2, 2.0] & [0.1, 2.2]\\
    $\log M_{\mathrm{cut}}$ & [10.0, 13.7] & [11.5, 14]\\
    $\sigma_{\log M}$ & [0.1, 1.6] & [0.08, 1.7]\\
    $v_{\mathrm{bc}}$ & [0, 0.7] & [0, 0.85]\\
    $v_{\mathrm{bs}}$ & [0.2, 2.0] & [0.1, 2.2]\\
    $c_{\mathrm{vir}}$ & [0.2, 2.0] & [0.1, 2.2]\\
    $\gamma_{f}$ & [0.5, 1.5] & [0.25, 1.75]\\
    $f_{\textrm{max}}$ & [0.1, 1] & [0.1, 1]\\
    \hline
\end{tabular}
\caption{All model parameters divided into cosmological and HOD parameters, with the training range used by the \textsc{Aemulus} emulator and the prior range used in the MCMC fit. Prior ranges were chosen to be slightly larger than the original training ranges, except where excluded by the physical meaning of the parameter, in order to be able to identify if the fit converges outside of the training range. The purpose of this extended range is only to more easily identify a prior dominated fit, since the emulator is not expected to produce accurate clustering outside of the training range. Instead, it would regress to the mean prediction. The exception is $\log M_{\rm cut}$, where the prior excludes the lower part of training range since $\log M_{\rm cut}$ ceases to have any impact on the halo occupation if it is below $\log M_{\rm min}$. This is the case for the eBOSS LRG sample, so $\log M_{\rm cut}$ is poorly constrained. However, we found the chains tended to pile up at the lower end of the training range, which gave the misleading impression that the data strongly preferred the lowest possible value, although it had no effect on the cosmological constraints. For that reason we set a more reasonable lower limit on $\log M_{\rm sat}$ for our sample.}
\label{table:parameters}
\end{table}

We assume our correlation function measurements are drawn from a multivariate Gaussian distribution, and use uniform priors for all model parameters, given in Table~\ref{table:parameters}. We explore the posterior surface for the fit between data and the \textsc{aemulus} correlation function predictions using a Markov-Chain Monte-Carlo (MCMC) sampler within the Cobaya \footnote{Cobaya, a \textbf{co}de for \textbf{bay}esian \textbf{a}nalysis in cosmology, is the Python successor to CosmoMC. Users are able to use the same MCMC sampler as CosmoMC \citep{Lewis:2002ah,Lewis:2013hha} in a Python framework, while allowing access to likelihoods from many major cosmological datasets. The sampler is tailored for parameter spaces with a speed hierarchy and implements the "fast dragging" procedure described in \citet{Neal:2005}. See https://cobaya.readthedocs.io for details.} framework \citep{torrado:2020xyz}. We include the full \textsc{Aemulus} HOD parameter space in our fit, however we limit the wCDM cosmological parameter space by fixing $N_{\rm eff}=3.046$ and $w=-1$, since these parameters are not well constrained by our measurements but have been well measured by other probes.

A concern for our small-scale analysis is that the separation range we use lacks a distinctive feature with a known scale to constrain the cosmological parameters, such as the BAO bump in large-scale analyses. Consequently, we consider a number of additional cosmological priors in order to set an accurate cosmology for our analyses. To begin with, we apply a uniform prior on the cosmological parameters based on the distance in 7D cosmological parameter space between the chain point and the cosmologies of the \textsc{aemulus} simulations used to train the emulator. If the distance is above a certain threshold the proposed step is forbidden, thus restricting the parameter space to the region which is well sampled by the training data, rather than the full uniform prior range given in Table~\ref{table:parameters}. In practice, the main impact of the training prior is to add the restriction $\sigma_8>0.65$, since there is only one training cosmology with $\sigma_8$ below that range. 

We also consider jointly fitting our data with the Planck 2018 TT,TE,EE and lensing likelihoods \citep{Aghanim:2019ame, Aghanim:2018oex} using the CAMB cosmological Boltzmann code \citep{Lewis:1999bs,Howlett:2012mh}, which constrain the cosmological parameters that control the shape of the power spectrum. It is important to note that $\gamma_f$ is treated as a free parameter in addition to the standard cosmological parameters, and is only constrained by RSD as measured from the eBOSS data. In effect, it represents a consistency check between the large-scale structure and CMB data: if these are consistent, we expect that $\gamma_f$ is close to one. We further consider three cases of the joint eBOSS and Planck fit. The first is a simple joint fit, where all of the cosmological parameters, including $\sigma_8$, and jointly fit by both the eBOSS clustering measurements through \textsc{aemulus} and the Planck likelihoods, while the HOD parameters and $\gamma_f$ are fit solely by the clustering measurements. The second is similar, except we explicitly account for the slight redshift offset between the emulator ($z=0.7$) and the data ($z=0.737$). The emulator takes all cosmological parameters at $z=0$, so the shape of the linear power spectrum will be identical between the cosmology described by the Planck likelihoods and the emulator, however there will be a difference in amplitude due to the slight redshift offset. Therefore, we adjust the value of $\sigma_8$ given to \textsc{aemulus} as follows
\begin{equation}  \label{eq:sig8_1}
    \sigma_{8,Aem} = \sigma_8(z=0) \times \frac{D(z=0.737)}{D(z=0)} 
    \times \frac{D(z=0)}{D(z=0.7)} \,.
\end{equation}
This makes sure that the normalisation of the \textsc{aemulus} output matches that expected at $z=0.737$ in the cosmology being tested: the first ratio corrects from $z=0$ to $z=0.737$ in the cosmology being tested, and the second ratio corrects from $z=0.7$ to $z=0$, where the normalisation is defined by \textsc{aemulus}. Thirdly, we consider a joint fit where the Planck likelihoods are used to constrain all of the cosmological parameters except for $\sigma_8$, which is fit solely by the clustering data. We test the robustness of our results to the inclusion of the training prior and the Planck likelihoods through these three methods in Sec.~\ref{sec:results_cosmo_priors}. Based on the results of these tests we use the training prior but not the Planck likelihoods for our fiducial analysis.

\section{Robustness and systematic error checks} \label{sec:model_tests}

In this section we explore the robustness of our model in general and to several possible sources of systematic error in particular. We begin by assessing the impact of non-linear velocities on our measurements, and what information is included from different scales. We then perform a general check of our method by fitting to measurements made on a mock catalogue. Finally, we check the impact of the two possible discrepancies between our model and the data, the effects of galaxy selection on the completeness of the HOD model, and redshift uncertainty.

\subsection{Contribution of non-linear velocities} \label{sec:non-linear}

\begin{figure*}
	\includegraphics[width=\textwidth]{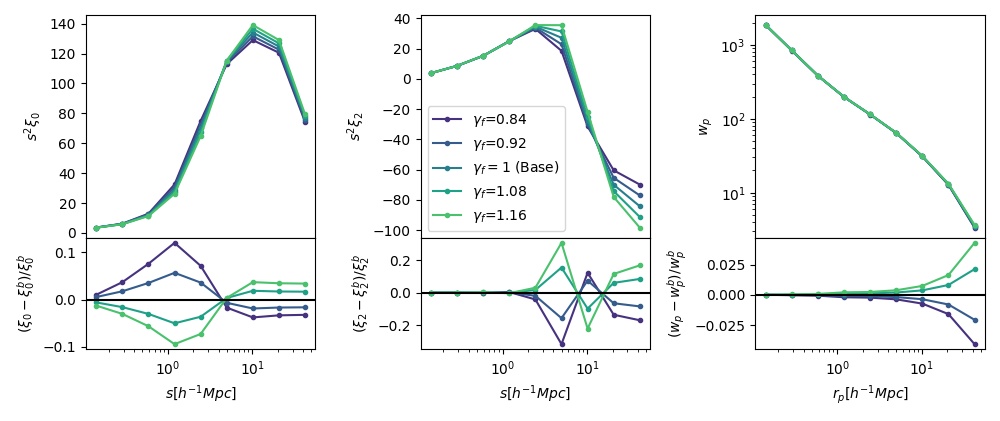}
    \caption{The effect on the emulator prediction of varying $\gamma_f$ for the monopole (\textit{left}), quadrupole (\textit{centre}), and projected correlation function (\textit{right}). All other parameters are kept fixed at reasonable values for the baseline eBOSS fit. \textit{Upper panels:} Direct comparison of the predictions, ranging from low $\gamma_f$ (blue) to high $\gamma_f$ (red). \textit{Lower panels:} Relative difference to the $\gamma_f=1$ prediction.}
    \label{fig:var_gamma_f}
\end{figure*}

In Sec.~\ref{sec:growth_meas} we introduced the key parameter of our measurement, $\gamma_f$, and described its significance on linear and non-linear scales. In order to identify the transition between these regimes we examine how the emulator prediction changes for various values of $\gamma_f$, shown in Fig.~\ref{fig:var_gamma_f}. For the three largest bins, varying $\gamma_f$ produces an almost constant relative change in the monopole, with a larger growth rate giving a larger clustering amplitude, as expected from linear theory. In the middle three bins the effect on the monopole changes signs as the quasi-linear regime transitions to the non-linear regime, where the random virial motions of the halos begin to dominate and increasing $\gamma_f$, which rescales all halo velocities, begins to damp the clustering. In the three smallest bins the effect of $\gamma_f$ on the monopole begins to decrease as the one-halo term begins to dominate. Because $\gamma_f$ affects only the halo velocities, and in our HOD formalism we do not assign galaxies based on subhalos, varying $\gamma_f$ has no effect on the one-halo term. Motivated by this result we divide our 9 measurement bins into three groups of three bins, with individual ranges of $0.1-0.8\mhmpc$, $0.8-7\mhmpc$, and $7-60\mhmpc$. These three ranges correspond roughly to the strongly non-linear regime where the one-halo term is dominant, the transition between the non-linear and quasi-linear regimes, and the quasi-linear regime. We therefore restrict our measurement of $f\sigma_8$ to the quasi-linear regime, where $\gamma_f$ can be interpreted as a rescaling of the linear growth rate. For measurements performed over the full separation range we instead use $\gamma_f$ as a test of $\Lambda$CDM, where a deviation from $\gamma_f=1$ indicates that the velocity field of the data as parameterized by our emulator model is in disagreement with the expectation from $\Lambda$CDM.

\subsection{Galaxy selection and the HOD model} \label{sec:fmax_test}

\begin{figure*}
	\includegraphics[width=\textwidth]{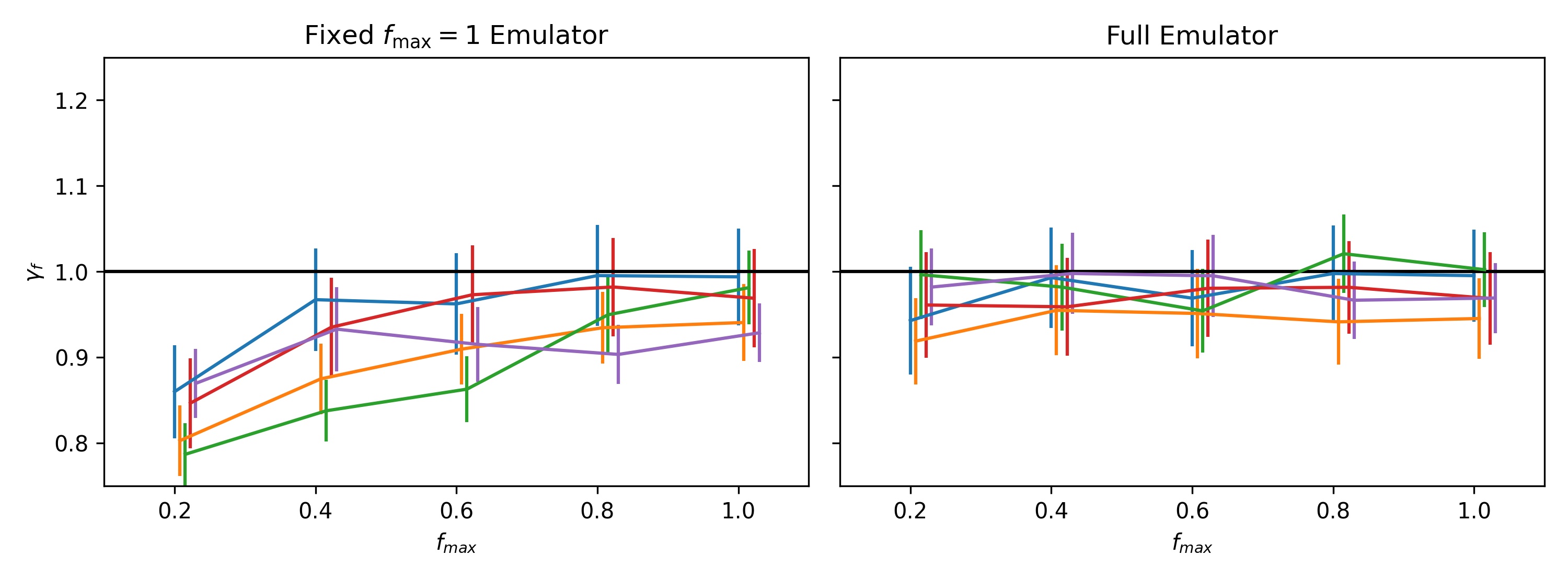}
    \caption{Performance of emulators with fixed or variable $f_{\rm max}$ on HOD mocks constructed with varying $f_{\rm max}$. The left panel shows the results from an emulator built with the original \textsc{Aemulus} parameter set, which is equivalent to $f_{\rm max}=1$. The right panel shows the results from the emulator used in our analysis with variable $f_{\rm max}$. Both emulators were built to match the eBOSS redshift and number density. The horizontal line shows the expected value of $\gamma_f$ used to construct the mocks. Points are shifted slightly along the x-axis to avoid overlap.}
    \label{fig:fmax_test}
\end{figure*}

As described in Sec.~\ref{sec:aemulus}, we add an additional parameter $f_{\rm max}$ to the emulator compared to previous uses that controls the maximum occupation fraction of central galaxies in the HOD framework, in order to address the incompleteness of the eBOSS LRG sample due to target selection. We test the necessity of this addition and the effect on the clustering using a series of HOD mock galaxy catalogues. We constructed these mocks from the Uchuu\footnote{http://skiesanduniverses.org/Simulations/Uchuu/} simulation. Briefly, Uchuu is a $(2000 \mhmpc)^3$, $12800^3$ particle simulation using the Planck2015 cosmology and a mass resolution of $m_p=3.27\times10^8 \,h^{-1}M_\odot$. We construct the mocks from the $z=0.7$ slice, using the \texttt{halotools}\footnote{https://halotools.readthedocs.io/en/latest/} \citep{halotools} Python package and a HOD parameterization identical to that outlined in Sec.~\ref{sec:aemulus}. We constructed mocks using $\sigma_{\log M}$, $\log M_{\mathrm{sat}}$, $\alpha$, and $\log M_{\mathrm{cut}}$ from five randomly selected test HOD parameter sets in \textsc{Aemulus}, with $\log M_{\mathrm{min}}$ tuned to give $n=1\times10^{-4}$. The \textsc{Aemulus} test HOD sets are themselves randomly selected from the uniform training range given in Table~\ref{table:parameters}, but were not used in training the emulator. In all mocks we kept the additional parameters $v_{\rm bc}=0$, $v_{\rm bs}=1$, $c_{\rm vir}=1$, and $\gamma_f=1$ fixed to their simplest, no scaling values. For each of the five HOD parameter sets we then constructed five mocks with $f_{\rm max}=[0.2, 0.4,0.6,0.8,1.0]$, for a total of 25 mocks.

We fit these 25 HOD mocks using two emulators: one matching the original \textsc{Aemulus} HOD model that is equivalent to fixing $f_{\rm max}=1$, and the full emulator with variable $f_{\rm max}$. Both emulators were built to match the eBOSS redshift and number density, as described in Sec.~\ref{sec:aemulus}. The $\gamma_f$ constraints on the HOD mocks from both emulators are shown in Fig.~\ref{fig:fmax_test}, where the expected value is $\gamma_f=1$ by the construction of the mocks. It should be noted that all of the mocks were constructed using the same halo catalog from a single simulation box at a particular cosmology, so it is unsurprising that the constraints do not scatter evenly above and below $\gamma_f=1$, since they are not fully independent. The key points to notice are that the variable $f_{\rm max}$ emulator is able to recover the expected value of $\gamma_f$ within the uncertainty over the full $f_{\rm max}$ range, and shows no trend in $f_{\rm max}$. Conversely, the fixed $f_{\rm max}$ emulator shows a clear bias in $\gamma_f$ for $f_{\rm max}\leq0.6$. This result matches what we would theoretically expect for model which overestimates the $f_{\rm max}$ value of the sample. If the mismatch is small there is not a significant change in the galaxy bias of the sample, however if $f_{\rm max}$ is significantly overestimated then the model prediction has a larger galaxy bias, $b$, then the sample, which is compensated by a lower growth rate since the amplitude of the linear clustering scales as $fb^2$.

\subsection{Redshift uncertainty} \label{sec:zerr_test}

\begin{figure}
	\includegraphics[width=\columnwidth]{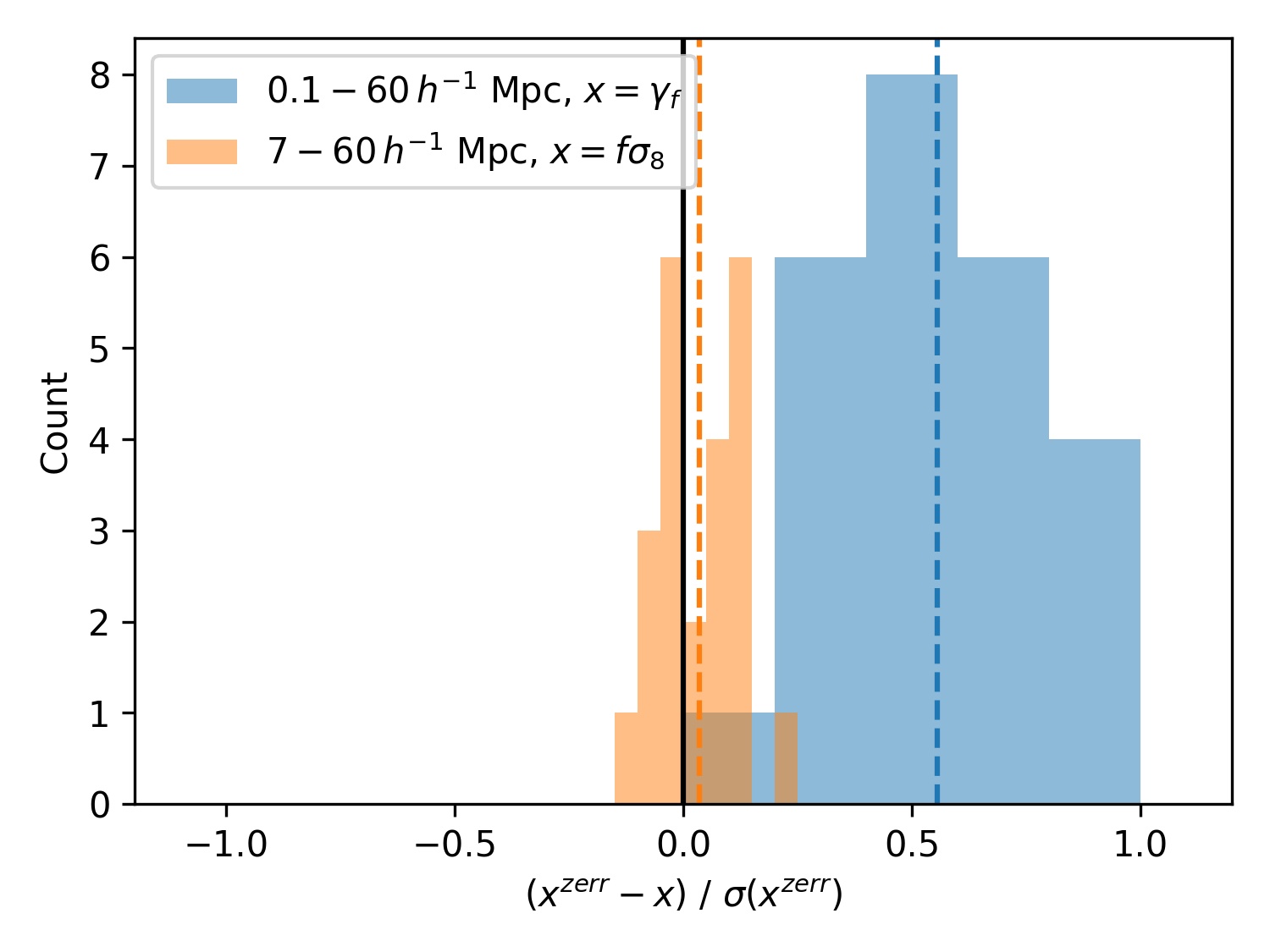}
    \caption{A histogram of the shifts in the measured cosmological parameters for 25 HOD mocks with and without a random velocity dispersion matching the eBOSS redshift uncertainty. Blue bars show the shift in $\gamma_f$ measured over the full separation range, and orange bars show the shift in $f\sigma_8$ measured from the quasi-linear scales only. The x-axis shows the difference between the value measured for the mock with a random velocity dispersion ($zerr$) and the value measured from the same mock without the additional velocity dispersion, divided by the uncertainty of the measurement from the $zerr$ mock. Coloured dashed lines show the mean shift for each case. For the fit over $0.1-60\mhmpc$, including a random velocity dispersion not represented in the model increased the measured value of $\gamma_f$ for all 25 mocks, with a mean shift slightly larger than half of the statistical uncertainty. Conversely, for the fit over $7-60\mhmpc$, the shifts from including a random velocity dispersion scatter around 0, with a mean shift that is negligible compared to the statistical error.}
    \label{fig:zerr_test}
\end{figure}

Another area of concern where the emulation based model may not accurately reflect the data is the effect of redshift uncertainties. As shown in Fig.2 of \citet{Ross20a}, the eBOSS LRG sample has a redshift uncertainty that is well approximated by a Gaussian with mean $\mu=1.3\rm~km~s^{-1}$ and standard deviation $\sigma=91.8\rm~km~s^{-1}$. On average, this means that each redshift is wrong by an absolute offset of $65.6 \rm~km~s^{-1}$. To first order this gives a Gaussian random velocity shift for all targets, which acts to damp the clustering of the multipoles on small scales. The parameters $v_{\rm bc}$ and $v_{\rm bs}$, which control the velocity dispersion of centrals and satellites respectively, should be able to mimic much of this effect in the model without affecting the constraints on other parameters. However, since $\gamma_f$ scales all halo velocities in the simulation, on non-linear scales where the halo velocities are virialized, $\gamma_f$ has a similar effect on the clustering as the redshift uncertainty, $v_{\rm bc}$ and $v_{\rm bs}$. In addition, $v_{\rm bc}$ and $v_{\rm bs}$ are both calculated by scaling the virial dispersion of the host halo, so the galaxy velocities derived in the model have a mass dependence which is not reflected in the redshift uncertainty. The result is that the redshift uncertainty may bias the recovered value of $\gamma_f$ on non-linear scales, with an unmodelled redshift uncertainty giving a larger than expected value of $\gamma_f$.

We test the effect of the redshift uncertainty on the $\gamma_f$ and $f\sigma_8$ constraints using a second set of HOD mocks, constructed in the same way as those described in Sec.~\ref{sec:fmax_test}. We selected 25 new \textsc{Aemulus} test HOD parameter sets and generated HOD catalogues using \texttt{halotools}. We then calculated the clustering with and without a random velocity shift along the line of sight drawn from a Gaussian with mean $\mu=1.3\rm~km~s^{-1}$ and standard deviation $\sigma=91.8\rm~km~s^{-1}$. The change in the measured values of $\gamma_f$ from the full separation range and $f\sigma_8$ from the quasi-linear scales only (matching the method used for our baseline results) due to the inclusion of the random velocity shift are shown in Fig.~\ref{fig:zerr_test}. For all 25 mocks, including a random velocity shift increased the value of $\gamma_f$ measured from the full separation range, with an average shift slightly greater than half the statistical uncertainty. The larger value of $\gamma_f$ measured due to the random velocity shift matches our theoretical expectation for the degeneracy between $\gamma_f$ and the redshift uncertainty on non-linear scales, and the magnitude of the shift indicates that the redshift uncertainty is a significant concern when fitting to the non-linear scales. On the other hand, the shifts in the measured value of $f\sigma_8$ scatter around 0, with a mean shift over an order of magnitude smaller than the statistical uncertainty. This result also agrees with what is expected for our model, since on quasi-linear scales the redshift uncertainty is not degenerate with a change in $\gamma_f$, and instead will change only $v_{\rm bc}$ and $v_{\rm bs}$. Therefore, the redshift uncertainty is not a concern for our value of $f\sigma_8$ measured from the quasi-linear scales.

There are several barriers to including a correction for the redshift uncertainty in the model. Most significantly, the redshift uncertainty grows with redshift (see Fig.6 of \citealt{Bolton:2012} for BOSS redshift evolution), while the emulator is constructed from catalogues at a single redshift slice. The evolution with redshift is also important because the eBOSS LRG targeting cuts were made using the apparent magnitudes of the targets, so properties of the sample such as the mean mass will also evolve weakly with redshift and correlate with the growth of the redshift uncertainty. The result is that including the redshift uncertainty in the model may not be as simple as drawing from a uniform velocity shift, and would require more detailed testing and corrections. The effect of redshift uncertainty could instead be included as an additional systematic error or shift in our measured values. However, it is important to note that for every mock tested the inclusion of redshift uncertainty (without it being present in the model) increased the measured value of $\gamma_f$, because on the non-linear scales where the redshift uncertainty is the most significant it is degenerate with the larger random motions of the halos provided by a larger value of $\gamma_f$. In Sec.~\ref{sec:results} we consistently measure values of $\gamma_f$ that are below the value expected from $\Lambda$CDM+Planck2018, so the presence of redshift uncertainty is actually expected to increase this tension rather than lowering it. We therefore take the conservative approach of excluding a shift in our measurements due to the redshift uncertainty, even though it would be expected to increase the tension shown by our measurements, and leave a complete treatment of the redshift uncertainty to future work.

\subsection{SHAM Mocks}

\begin{figure}
	\includegraphics[width=\columnwidth]{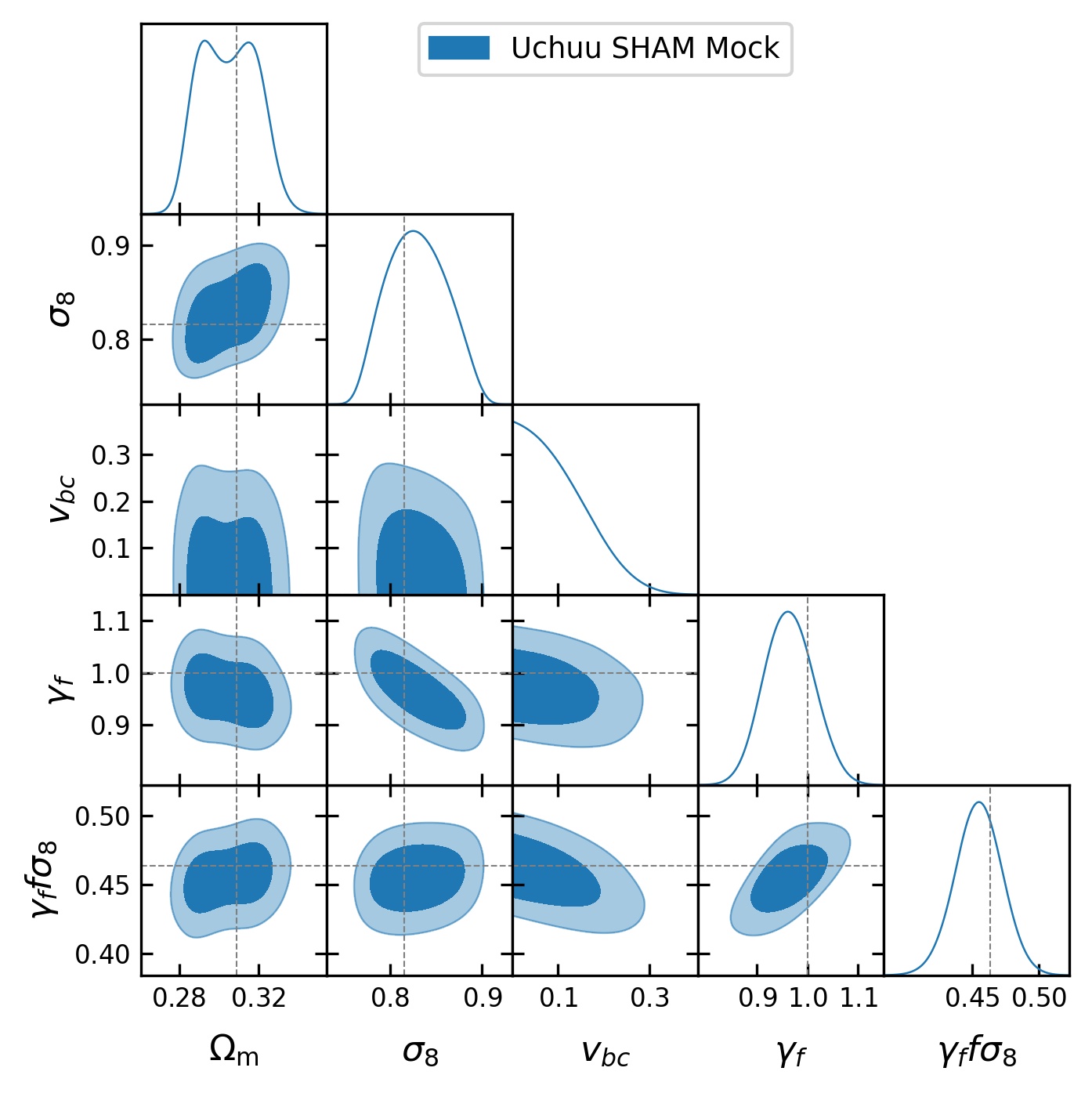}
    \caption{2D and 1D marginalized constraints of the key parameters from the fit to an Uchuu SHAM mock matching the eBOSS LRG number density and redshift. Dotted lines show the values of the cosmological parameters from the simulation.}
    \label{fig:sham-tp}
\end{figure}

We test the robustness of our model and analysis pipeline using a subhalo abundance matching (SHAM) mock generated from the Uchuu simulation. By using a SHAM mock rather than a HOD mock we remove the dependence on the specific galaxy-halo connection model used in our analysis, providing the best approximation to a model independent test. If our analysis is able to correctly recover the expected value of $\gamma_f=1$ for the SHAM mock then we can be confident it will be able to match the data, even if there are deviations from the specific functional form of the galaxy-halo connection model described in Sec.~\ref{sec:aemulus}. We use the $z=0.7$ slice of the simulation to construct a SHAM mock using the peak halo velocity, $V_{\rm peak}$, with a scatter of 0.2 dex, and a number density of $n=1\times10^{-4}$ in order to match the eBOSS LRG number density and redshift.

The result of our fit to the SHAM mock is shown in Fig.~\ref{fig:sham-tp}. The primary purpose of the Uchuu SHAM mock test is to assess the robustness of the cosmological parameter recovery using our HOD based emulator, so we have only included the parameters which have the greatest impact on the $\gamma_f$ constraint. The constraints on all of the cosmological parameters are in good agreement with the known values from the simulation, and the 1D marginalized constraint on $\gamma_f$ is $\gamma_f=0.964\pm0.049$, which agrees to within $1-\sigma$ with the known value of $\gamma_f=1$ for the mock. All well constrained HOD parameters converge within the training parameter space indicating that the emulator is able to accurately model the clustering of the mock, despite the mock being constructed using a different galaxy-halo connection. This result shows that are analysis pipeline and model provide robust constraints on the growth rate.

\section{Results}  \label{sec:results}

In this section we present the results of our fit to the small-scale LRG clustering. We also investigate the robustness of our results by testing the inclusion of additional constraints on the cosmological parameters, examining how the constraints change depending on which scales and measurements are included in the analysis, the effect of covariance matrix smoothing on the measured parameters, and consistency with the constraints from a combined CMASS+eBOSS sample.

\subsection{Headline results} \label{sec:baseline_results}

\begin{figure*}
	\includegraphics[width=\textwidth]{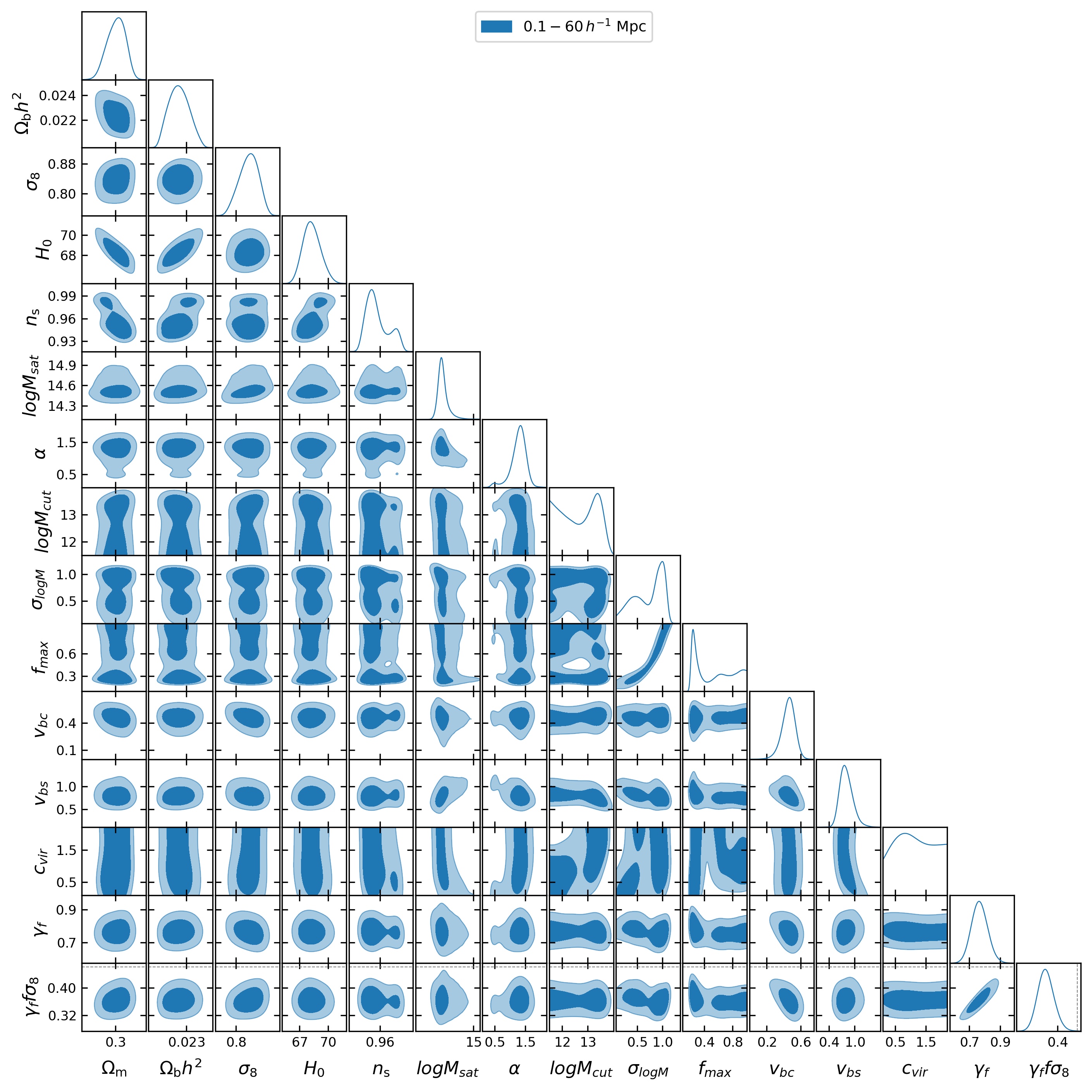}
    \caption{1D and 2D contours of the parameters used in our baseline fit, as well as the derived constraints on $\gamma_f f\sigma_8$.}
    \label{fig:triangle-plot}
\end{figure*}

\begin{figure*}
	\includegraphics[width=\textwidth]{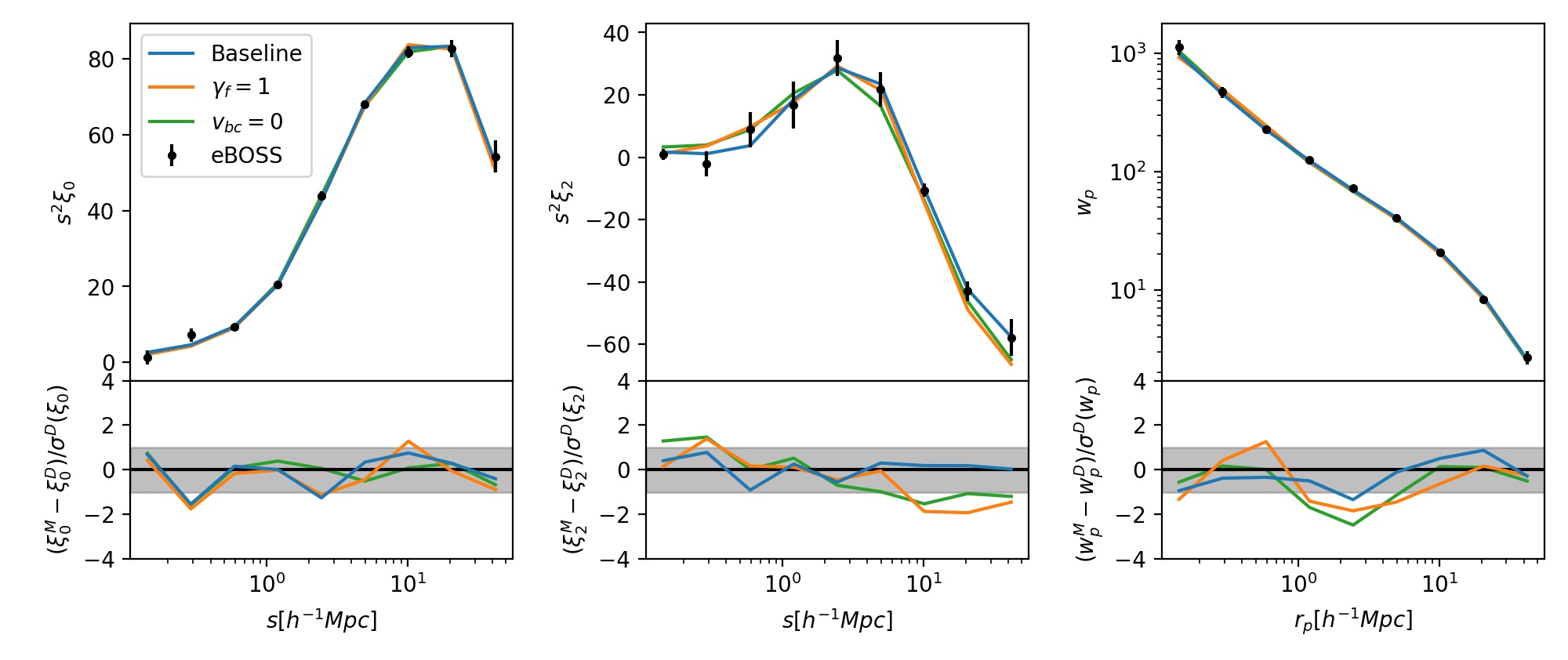}
    \caption{Comparison of the best fit model predictions to the data several fit to the eBOSS LRG sample for the monopole (\textit{left}), quadrupole (\textit{centre}) and projected correlation function (\textit{right}). \textit{Upper panels:} The baseline fit (blue), fixed $\gamma_f=1$ fit (orange), and $v_{\rm bc}=0$ fit (green), with the data and measurement uncertainty (black). \textit{Lower panels:} The difference between the best fit models and the data divided by the measurement uncertainty. The $1-\sigma$ region is shown in grey.}
    \label{fig:model-comp}
\end{figure*}

We fit the eBOSS LRG monopole, quadrupole, and projected correlation function over scales $0.1<r<60\mhmpc$ using the Cobaya MCMC sampler. We restrict the cosmological parameter space using the \textsc{Aemulus} training prior described in Sec.~\ref{sec:fit}, but do not include any external data. We obtain a value of $\gamma_f=0.767\pm0.052$, $4.5\sigma$ below what would be expected in a $\Lambda$CDM+GR universe. The 1D and 2D likelihood contours of the full parameter set are shown in Fig.~\ref{fig:triangle-plot}. All well constrained parameters are within the prior ranges described in Table~\ref{table:parameters}, and the parameters that are most impactful for our results, $\Omega_m$, $\sigma_8$, $v_{\rm bc}$, and $\gamma_f$, all show roughly Gaussian constraints. The best-fit values of the cosmological parameters other than $\gamma_f$ are consistent with recent measurements from the Planck Collaboration \citep{Planck-2018-params}. The best fit model prediction is plotted relative to the data in Fig.~\ref{fig:model-comp}, showing reasonable agreement within the measurement uncertainty on all scales. The best fit prediction has $\chi^2=14.1$, with 14 degrees of freedom and 27 data points, indicating a good fit. In addition, we consider a fit over only the quasi-linear scales of our measurements, $7-60\mhmpc$ as described in Sec.~\ref{sec:non-linear}, from which we obtain a value of $f\sigma_8(z=0.737)=0.408\pm0.038$. This value is $1.4\sigma$ below what is expected from the 2018 Planck data for a flat $\Lambda$CDM universe, and is a factor of 1.7 improvement in statistical error over the more standard large-scale analysis of the same data set. See Sec.~\ref{sec:scales_results} for more details.

\subsection{Testing the quasi-linear scales for overfitting} \label{sec:results_N_params
}

One concern for our fit to the quasi-linear scales is that by reducing the separation range to $7-60\mhmpc$ we are fitting nine data points with a 14 free parameter model. However, it is important to note that many of the HOD parameters have a negligible effect on these scales. In particular, the three parameters that control the satellite occupation ($\log M_{\rm sat}$, $\alpha$, $\log M_{\rm cut}$) and the three parameters that control the positions of galaxies in the halos ($v_{\rm bc}$, $v_{\rm bs}$, $c_{\rm vir}$) have very little impact and are almost entirely constrained by the $0.1-7\mhmpc$ bins. Therefore, while there are 14 free parameters in the model, only eight are significant when fitting to the nine bins of the quasi-linear scales. While this provides a theoretical explanation for why the quasi-linear scales will not be overfit, our fit over the scales $7-60\mhmpc$ has a minimum $\chi^2=0.36$ (Table.\ref{table:runs}), indicating that the small scale HOD parameters may still be causing some overfitting.

\begin{figure}
	\includegraphics[width=\columnwidth]{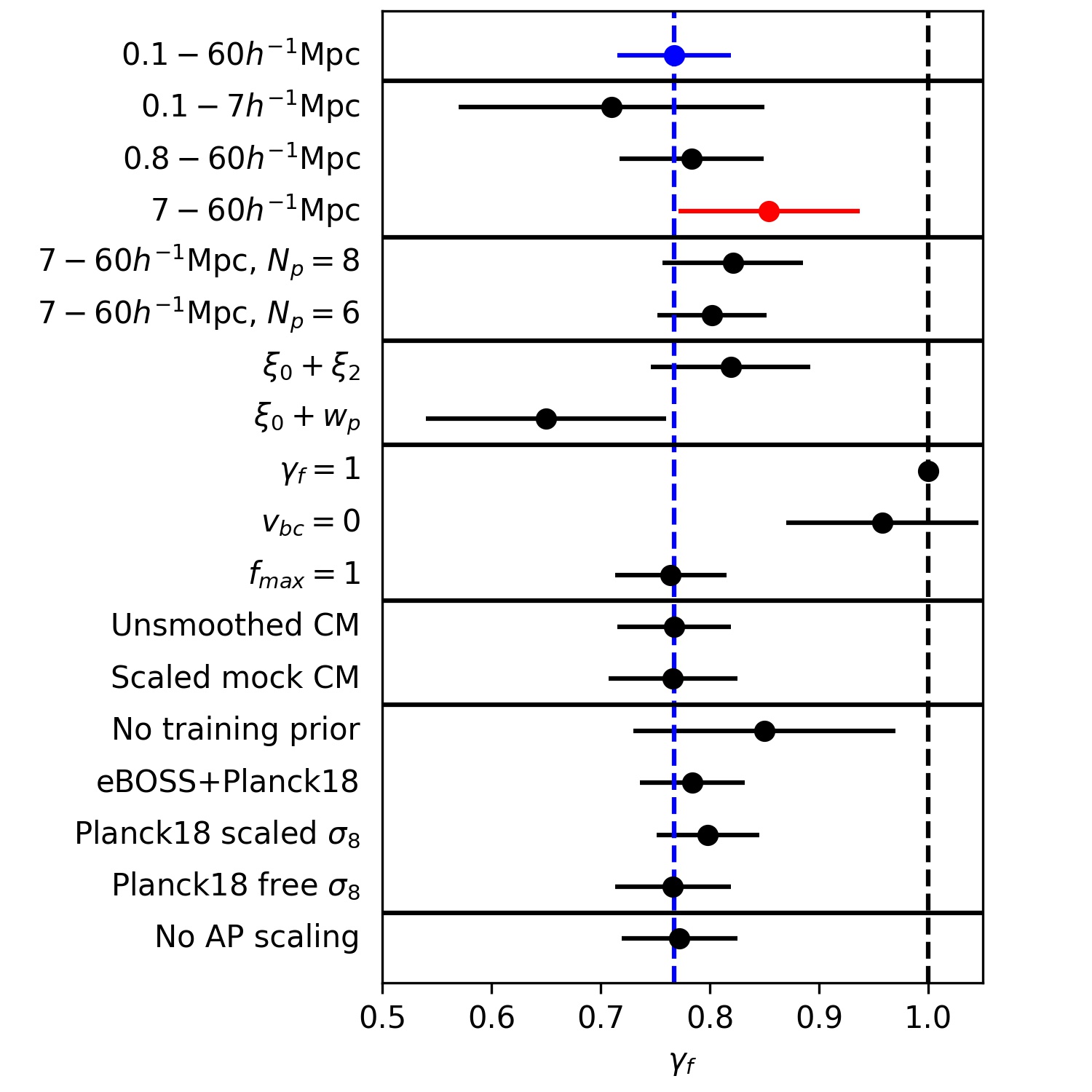}
    \caption{$\gamma_f$ constraints from all the runs listed in Table~\ref{table:runs}. The blue point shows the baseline fit to the full separation range, extended by the blue dashed line for comparison to other points. The red point shows the fit to the quasi-linear scales only. The black dashed line shows $\gamma_f=1$ for comparison, the value expected if the amplitude of the halo velocity field matches the $\Lambda$CDM expectation.}
    \label{fig:runs}
\end{figure}

To test if this overfitting affects our results we perform additional fits over the $7-60\mhmpc$ separation range with the predominantly small scale HOD parameters fixed to their best fit values from the fit over the full $0.1-60\mhmpc$ separation range. In the first additional fit we keep the six parameters listed above fixed, leaving eight parameters ($\Omega_m$, $\Omega_b h^2$, $\sigma_8$, $h$, $n_s$, $\sigma_{\log M}$, $\gamma_f$, $f_{\rm max}$) free. In the second fit we also keep $\sigma_{\log M}$ and $f_{\rm max}$ fixed to their best fit values from the full fit, allowing only the six cosmological parameters to vary. The $\gamma_f$ constraints from these fits are shown in Table.~\ref{table:runs} and Fig.~\ref{fig:runs}. The results of both fits show that reducing the parameter space increases the precision of the $\gamma_f$ constraint without significantly shifting the central value, while increasing the minimum $\chi^2$. We conclude that allowing the small scale HOD parameters to be free does lead to the quasi-linear scales being overfit, however it does not bias our cosmological constraints and instead only increases the uncertainty. Fixing these HOD parameters would increase the precision of our measurement from the quasi-linear scales, but it would also introduce an indirect dependence on the non-linear scales. We therefore take the conservative choice of using the measurement with all 14 parameters free as our baseline result. However, this test does show the value of including the non-linear scales in a measurement of the linear growth rate.

\begin{table}
\centering
\begin{tabular}{|l|c|c|c|c|}
    \hline
    Run & $\gamma_f$ & $N_P$ & $N_D$ & $\chi^2$\\
    \hline
    $0.1-60\mhmpc$ & $0.767\pm0.052$ & 14 & 27 & 14.1 \\
    \hline
    $0.1-7\mhmpc$ & $0.71\pm0.14$ & 14 & 18 & 7.8 \\
    $0.8-60\mhmpc$ & $0.783\pm0.066$ & 14 & 18 & 4.2 \\
    $7-60\mhmpc$ & $0.854\pm0.083$ & 14 & 9 & 0.36 \\
    \hline
    $7-60\mhmpc$, 8 parameters & $0.821\pm0.064$ & 8 & 9 & 0.74 \\
    $7-60\mhmpc$, 6 parameters & $0.802\pm0.050$ & 6 & 9 & 1.8 \\
    \hline
    $\xi_0+\xi_2$ & $0.819\pm0.073$ & 14 & 18 & 5.0 \\
    $\xi_0+w_p$ & $0.65\pm0.11$ & 14 & 18 & 5.4 \\
    \hline
    $\gamma_f=1$ & $1$ & 13 & 27 & 28.0 \\
    $v_{\textrm{bc}}=0$ & $0.958\pm0.088$ & 13 & 27 & 22.5 \\
    $f_{\textrm{max}}=1$ & $0.764\pm0.051$ & 13 & 27 & 16.6 \\
    \hline
    Unsmoothed covariance matrix & $0.767\pm0.052$ & 14 & 27 & 14.3 \\
    Scaled mock covariance matrix & $0.766\pm0.059$ & 14 & 27 & 12.0 \\
    \hline
    No training prior & $0.85\pm0.12$ & 14 & 27 & 12.1 \\
    eBOSS+Planck18 & $0.784\pm0.048$ & 14* & 27 & 18.5 \\
    eBOSS+Planck18 scaled $\sigma_8$ & $0.798\pm0.047$ & 14* & 27 & 19.1 \\
    eBOSS+Planck18 free $\sigma_8$ & $0.766\pm0.053$ & 14* & 27 & 18.0 \\
    \hline
    No AP scaling & $0.772\pm0.053$ & 14 & 27 & 14.5 \\
    \hline
\end{tabular}
\caption{$\gamma_f$ constraints with statistical errors calculated from the width of the 1D marginalized posterior and $\chi^2$ values for the fits used in our analysis. $N_P$ gives the number of free model parameters in the fit and $N_D$ gives the number of data points. *The eBOSS+Planck18 runs jointly fit 5 of the 14 parameters with Planck, so they are not fully independent.}
\label{table:runs}
\end{table}

\subsection{Testing the impact of the cosmological priors} \label{sec:results_cosmo_priors}

\begin{figure}
	\includegraphics[width=\columnwidth]{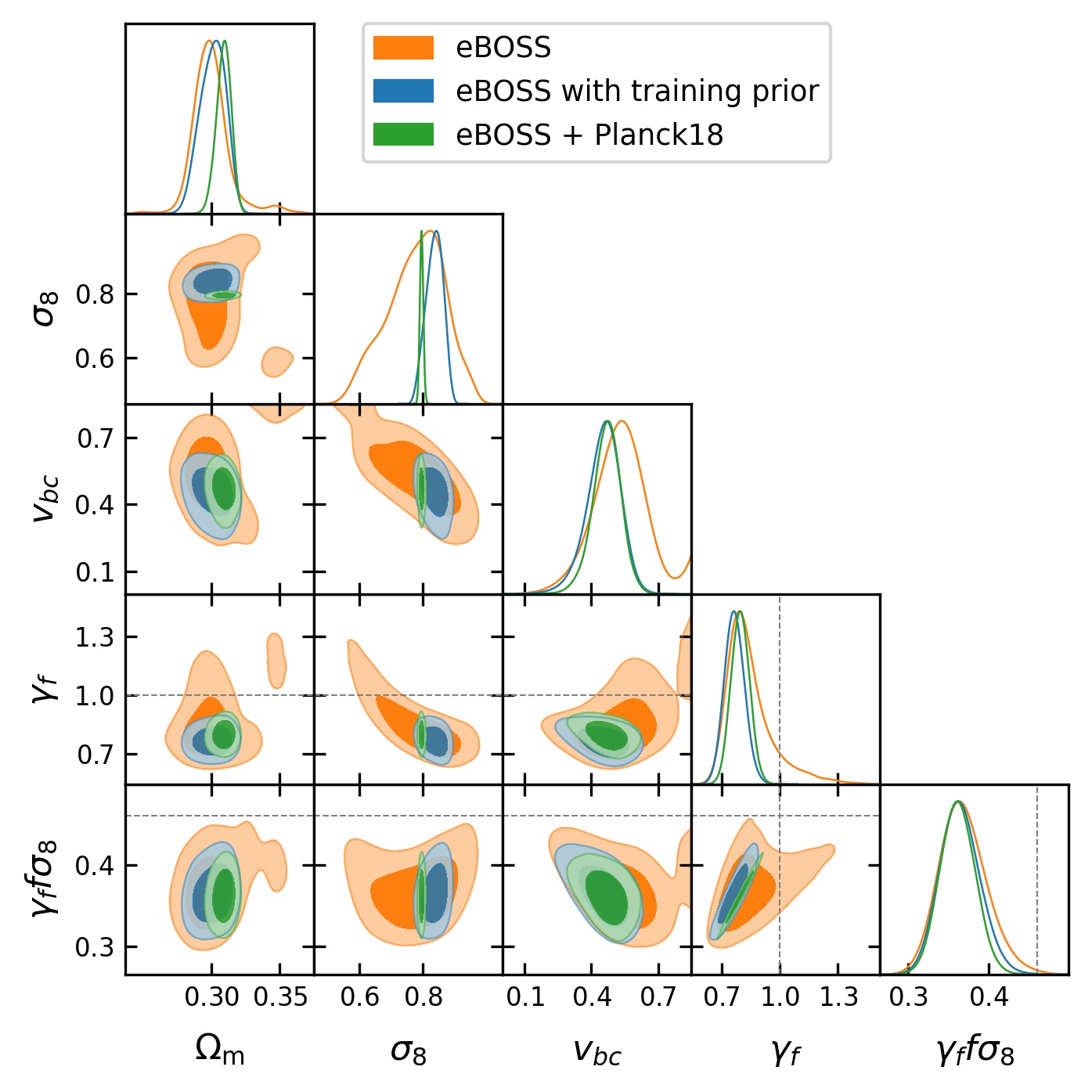}
    \caption{1D and 2D contours of the key fit parameters for the fit to the eBOSS LRG sample with no additional cosmological constraints (orange), restricted by the \textsc{Aemulus} training prior (blue), and jointly fit with the Planck2018 likelihoods (green).}
    \label{fig:ao-tp}
\end{figure}

We consider a number of prior constraints on the cosmological parameters, as described in Sec.~\ref{sec:fit}. The three most significant cases are a uniform prior as described in Table~\ref{table:parameters}, a uniform prior that restricts the cosmological parameters to be within the volume that is well sampled by the training simulations, and a joint fit with Planck2018 likelihoods with a scaled value of $\sigma_8$ to account for the redshift difference between the data and the model. The constraints on the key parameters for these three prior choices are shown in Fig.~\ref{fig:ao-tp}. The parameter that is most significantly impacted by the prior choice is $\sigma_8$, with all three methods giving consistent values but with large differences in precision. However, the constraint on $f\sigma_8$ is almost unchanged for all prior choices. This result clearly shows the robustness of the $f\sigma_8$ fit from the data, and demonstrates the freedom of the model where changes in $\sigma_8$ can be balanced by $\gamma_f$. It is also important to note that because the uncertainty on $f\sigma_8$ is dominated by the uncertainty of $\gamma_f$ that the training prior and the joint fit with Planck achieve almost the same precision on $f\sigma_8$, despite having comparable constraints on $\gamma_f$ but a significant difference in precision on $\sigma_8$.

The effect of the three treatments of $\sigma_8$ for the joint Planck fit described in Sec.~\ref{sec:fit} can be found in Table~\ref{table:runs}. Using the same value of $\sigma_8$ for the Planck chains and model, scaling to account for the redshift offset, or excluding the Planck constraints on $\sigma_8$ all give consistent values for the growth rate, again demonstrating the robustness of the fit.

\subsection{Testing the dependence on the data fitted} \label{sec:scales_results}

\begin{figure*}
    \centering
    \begin{subfigure}[b]{\columnwidth}
    	\includegraphics[width=\columnwidth]{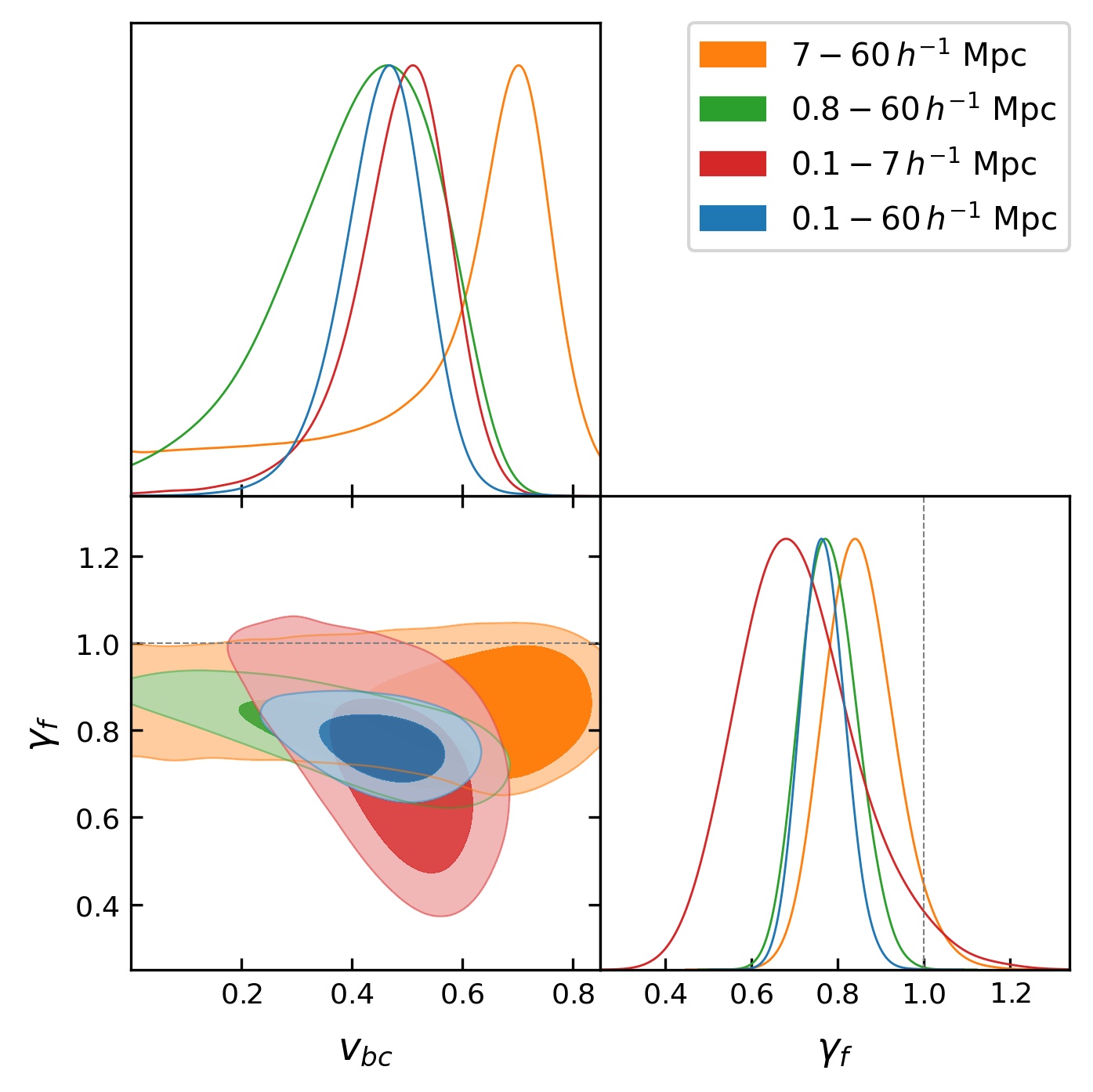}
        \label{fig:scales}
    \end{subfigure}
    \hfill
    \begin{subfigure}[b]{\columnwidth}
    	\includegraphics[width=\columnwidth]{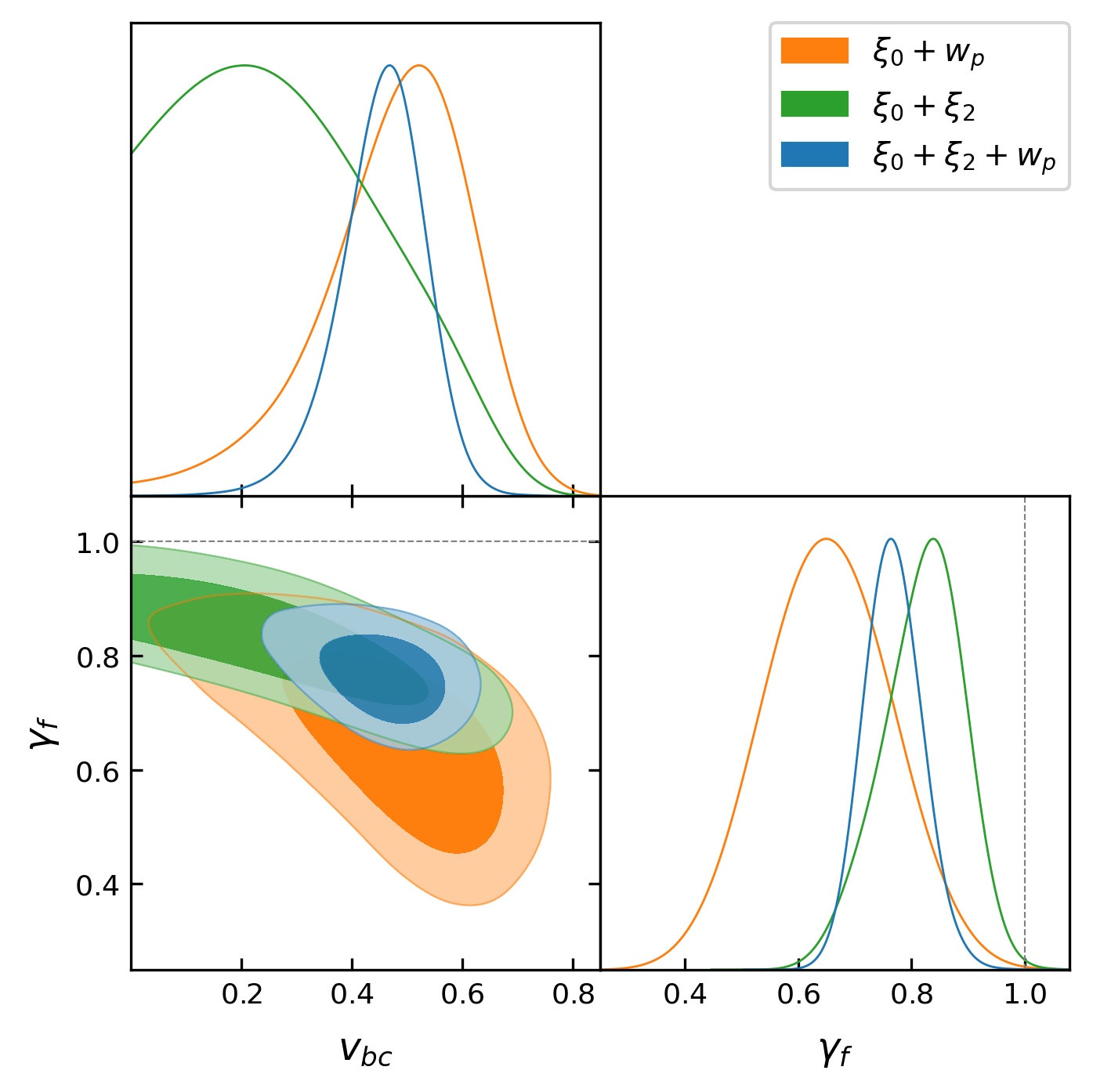}
        \label{fig:probes}
    \end{subfigure}
        \caption{2D and 1D marginalized constraints on $v_{\rm bc}$ and $\gamma_f$ for fits to different scales and measurements. \textit{Left:} Constraints from the three largest separation bins (orange), six largest separation bins (green), six smallest separation bins (red), and all nine separation bins (blue) for all three measurements. The dotted line shows $\gamma_f=1$, the value expected if the amplitude of the halo velocity field matches the expectation from $\Lambda$CDM. \textit{Right:} Constraints from the joint fit to the monopole and projected correlation function (orange), monopole and quadrupole (green), and all three measurements (blue).}
        \label{fig:scales-probes}
\end{figure*}

In order to test the consistency of the constraint on $\gamma_f$ from the different regimes described in Sec.~\ref{sec:non-linear} we fit to the full non-linear regime ($0.1-7\mhmpc$), the weakly non-linear and quasi-linear regimes ($0.8-60\mhmpc$), and the quasi-linear regime only ($7-60\mhmpc$). 1D and 2D contours in the $v_{\rm bc}-\gamma_f$ parameter space for these three fits are shown in the left panel of Fig.~\ref{fig:scales-probes}. There is little variation in the other parameters between these fits to different scales, however some important insight is gained from examining the $v_{\rm bc}-\gamma_f$ degeneracy since both parameters have a similar effect on the clustering in the non-linear regime. The fits to smaller scales yield larger and more precise values of $v_{\rm bc}$, while obtaining smaller and less precise constraints on $\gamma_f$. The full fit to all scales is located at the intersection in $v_{\rm bc}-\gamma_f$ space of the small and larger scale fits. The result is that there is mild tension between the constraints on small and large scales, although the significance when considering the combined uncertainty is less than $1-\sigma$. It is worth recalling that since $\gamma_f$ rescales all halo velocities in the simulation, in the linear regime it can be used to derive a constraint on the linear growth rate $f\sigma_8$, in the non-linear it also enhances the effects of non-linear growth. So the fit to the small-scales is really a consistency check between the data and model with $\Lambda$CDM, and these results showing that there is a strong tension which is most significant in the non-linear regime.

The fit to the quasi-linear scales only does not show the same degeneracy between $v_{\rm bc}$ and $\gamma_f$ since they no longer have the same effect on the clustering, and is broadly consistent with any value of $v_{\rm bc}$ since it ceases to be impactful on such large scales. However, the large scale fit is still able to recover a relatively tight constraint on $\gamma_f$ that can be compared directly to the linear growth rate, giving a measurement $f\sigma_8=0.408\pm0.038$, which is $1.4\sigma$ lower than the value expected from the 2018 Planck data for a flat $\Lambda$CDM model.

We also examine the effect of excluding certain measurements from the fit. In the right panel of Fig.~\ref{fig:scales-probes} we show the constraints in $v_{\rm bc}-\gamma_f$ parameter space from the joint fit to only the monopole and projected correlation function, and the joint fit to the multipoles only. The multipole only fit is less sensitive to the degeneracy between $v_{\rm bc}$ and $\gamma_f$, but prefers a smaller value of $v_{\rm bc}$ and larger $\gamma_f$ compared to the full fit. On the other hand the joint fit of the monopole and projected correlation function, which contain similar clustering information but are sensitive and insensitive to the effects of RSD respectively, prefer a non-zero value of $v_{\rm bc}$ with much greater confidence, compensated by a low but less well constrained value of $\gamma_f$. As with the fits to different scales, the full fit lies in the overlap region produced by the different sensitivities of these measurements.

\subsection{Testing the dependence on the covariance matrix}

We test the robustness of our covariance matrix smoothing by fitting to the unsmoothed covariance matrix and a scaled version of the covariance matrix estimated from 1000 EZmocks. These mocks are constructed to match the clustering of the eBOSS DR16 samples on mildly non-linear and linear scales, but are not matched on small-scales, where the mocks exhibit very different clustering from the data. EZmocks are based on a Gaussian random field in a $5\mhgpcc$ box and an initial power spectrum and geometry of a flat $\Lambda$CDM cosmology with parameters $\Omega_m=0.307115$, $\Omega_b=0.048206$, $h=0.6777$, $\sigma_8=0.8225$ , $n_s=0.9611$. Matter particles are displaced from their initial to final positions using the Zel'dovich approximation. Tracer bias relation is calibrated to match the observed clustering of the target sample in the eBOSS DR16 data. The linear component of the redshift-space distortions is imprinted using Zel’dovich approximation while the non-linear term is modelled through an isotropic Gaussian motion. Mocks are then trimmed to match the geometry and radial selection function of the eBOSS DR16 LRG catalogue. The unscaled mock covariance matrix displays a similar correlation structure to the covariance matrix calculated form applying jackknife to the data, however because the clustering of the mocks on scales below $\sim1\mhmpc$ ~is significantly lower than the data the mock covariance matrix underestimates the variance on those scales. To bring the mock covariance matrix into better agreement we calculate the correlation matrix from the mocks, and then convert the correlation matrix to the covariance matrix by scaling the original diagonal values of the mock covariance matrix according to

\begin{equation}  \label{eq:mock_scale}
    \sigma^{M,s}_{i,i} = \sigma^{M}_{i,i}\frac{\xi^D_i}{\bar{\xi}^M_i},
\end{equation}
 
 where $\xi^D$ is the correlation function from the data and $\bar{\xi}^M$ is the mean correlation function from the 1000 EZmocks. This scaling preserves the original correlation structure and $\sigma(\xi)/\xi$ ratio of the mock covariance matrix, but adjusts for the higher clustering of the data. However, this method results in a very large variance for the quadrupole bins because the the mean quadrupole of the mocks goes to 0 on small scales. In order to prevent this artificial inflation of the quadrupole bins we instead use $\sigma^{M,s}_{i,i} = \sigma^{D}_{i,i}$ for the quadrupole.

The results of the fits using this scaled mock covariance matrix and the original unsmoothed jackknife covariance are shown in Table~\ref{table:runs}. The constraints in both cases are nearly identical to our baseline fit using the smoothed jackknife covariance matrix, indicating that our analysis is robust to the choice of covariance matrix.

\subsection{Testing the dependence on AP correction}\label{sec:ap_results}

We test the dependence of our result on the AP correction by running a full fit excluding the AP correction. The constraint on $\gamma_f$ from this fit can be seen in Table~\ref{table:runs} and Fig.~\ref{fig:runs}. Excluding the AP correction has a negligible effect on the constraint on $\gamma_f$ and slightly increases the best fit $\chi^2$. We therefore conclude that any uncertainty in the AP correction due to the large bin width and approximate calculation will not have a significant effect on our cosmological constraints.

\subsection{Including the BOSS CMASS data} \label{sec:cmass_eboss_results}

\begin{figure*}
    \centering
    \begin{subfigure}[b]{\columnwidth}
    	\includegraphics[width=\columnwidth]{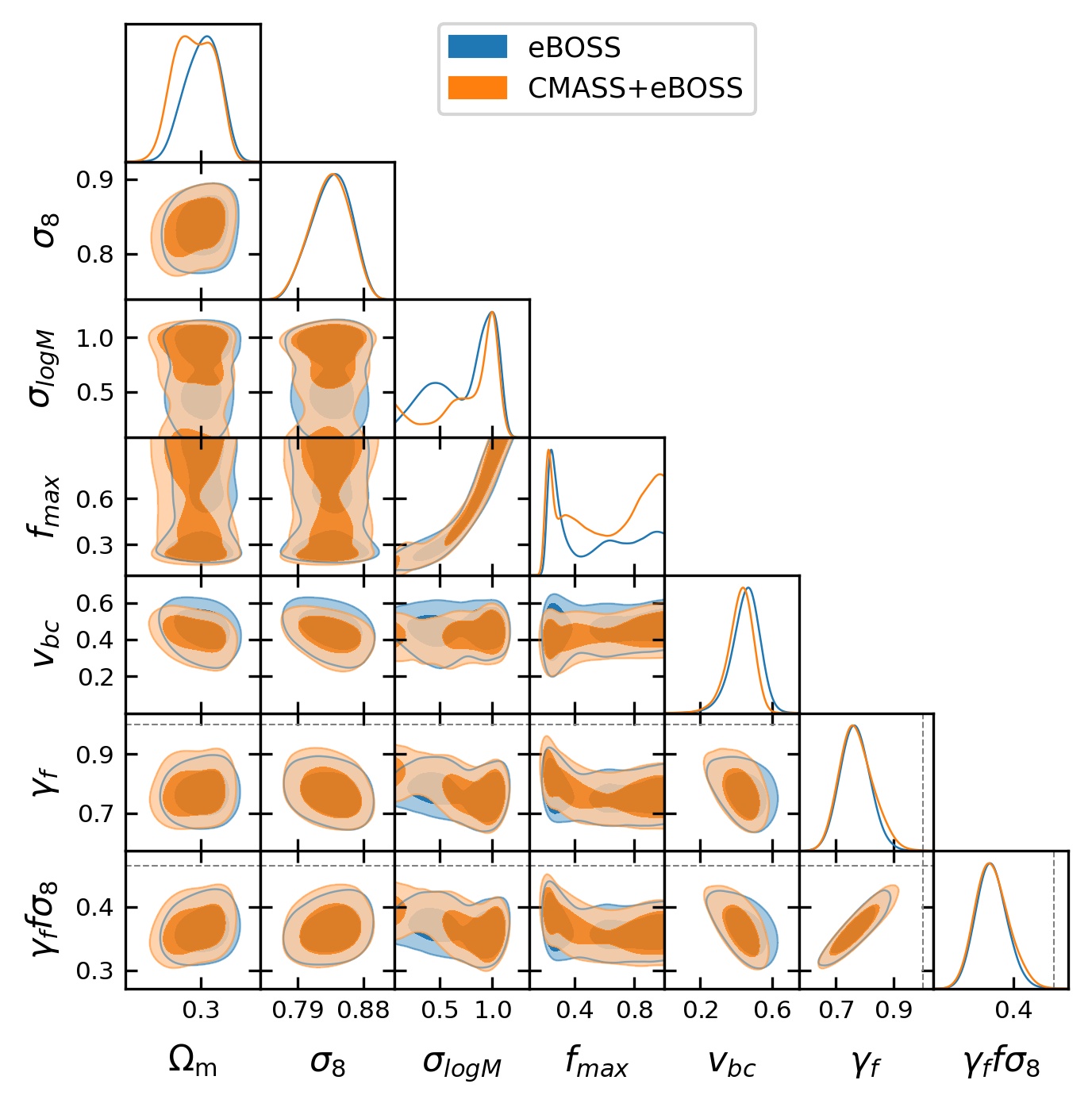}
    	\caption{$0.1-60\mhmpc$}
        \label{fig:cmass_eboss_full}
    \end{subfigure}
    \hfill
    \begin{subfigure}[b]{\columnwidth}
    	\includegraphics[width=\columnwidth]{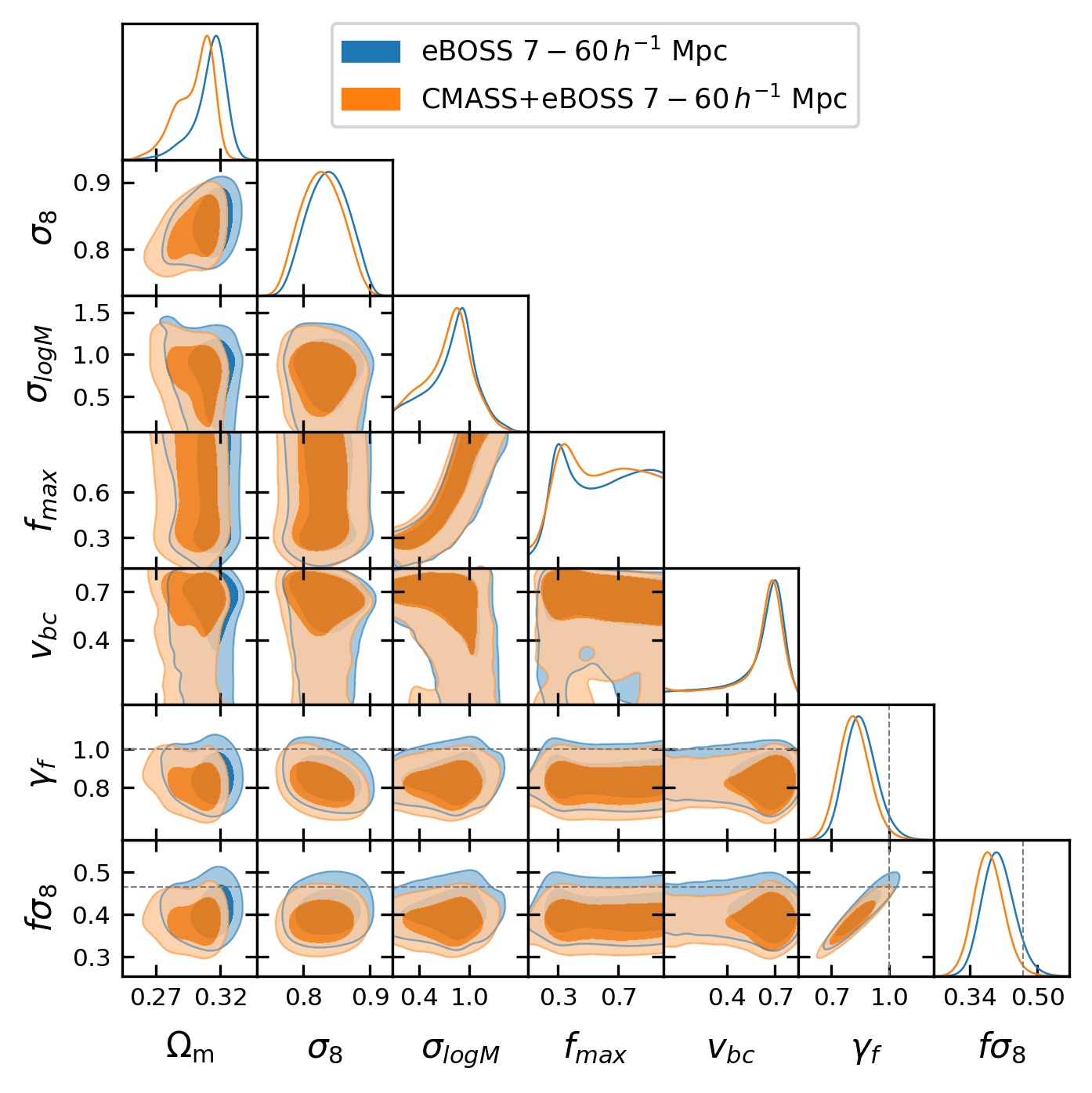}
    	\caption{$7-60\mhmpc$}
        \label{fig:cmass_eboss_large}
    \end{subfigure}
         \caption{2D and 1D marginalized constraints of the key parameters of our fit for our fiducial eBOSS measurement (blue) and combined CMASS+eBOSS sample (orange). The two plots show (a) the fit over the full emulator range, and (b) the fit to the quasi-linear scales only.}
        \label{fig:cmass_eboss}
\end{figure*}

We test the reliability of our fit using a combined CMASS+eBOSS sample in the redshift range $0.6\leq z \leq0.8$. In particular, in our analysis we use the CMASS sample from the DR12 data release. The CMASS DR12 catalogue covers an area of $9376\deg^2$ over a redshift range of $0.4<z<0.8$ \citep{Reid16} with a target density of $99.5\deg^{-2}$. The target selection is calibrated to provide a sample of galaxies with approximately constant stellar mass over the spanned redshift range. We refer the reader to \cite{Reid16} for a detailed description of the target selection and properties for CMASS sample. In order to perform a joint measurement of the two-point correlation function using the eBOSS and CMASS catalogues we restrict the two samples (and the corresponding random catalogues) only to the area of the sky where they overlap and to the redshift range of $0.6<z<0.8$. The redshift distributions of the two samples as well as their joint distribution are shown in Fig. \ref{fig:cmass_eboss_nz}.

The advantage of this sample is that it is more complete due to the complimentary nature of the CMASS and eBOSS colour cuts. However, the inclusion of the additional CMASS objects skews the redshift distribution of the sample, which is not ideal for an HOD-based analysis where the galaxy-halo connection parameters are implicitly assumed to be the same across the full redshift range of the sample, and several are dependent on the density of galaxies. As such we use our combined CMASS+eBOSS measurement to provide a consistency check with our fit, particularly our assumption that the target selection of eBOSS does not affect our measurement, but we continue to use the eBOSS only constraint as our fiducial measurement.

To correct fibre-collisions in the CMASS sample we use a modified version of the nearest-neighbour upweighting with completeness correction, designated CP, as described in Sec. 2.3 of \citet{Mohammad:2020}, and the standard angular upweighting method described in Sec.~\ref{sec:pip}. For the eBOSS LRG sample the CP correction was found to perform similarly to the PIP only result on all scales of $w_p$, $\xi_0$, and $\xi_2$ \citep[see Fig. 15 and 18 of ][]{Mohammad:2020}. Given the similarities in sample type and targeting between CMASS and eBOSS it is reasonable to expect a similar result for CMASS. When combined with angular upweighting any systematic bias is expected to be below the statistical uncertainty of the measurement. Since our primary goal in analyzing the combined CMASS+eBOSS sample is as a consistency check, this correction is sufficient for our purposes.

Fig.~\ref{fig:cmass_eboss} shows the result of our fit compared to the eBOSS only fit in the most important parameters of our analysis for both the full emulator range and the quasi-linear scales only. The CMASS+eBOSS measurement is consistent with the eBOSS only measurement in all parameters, although there is a greater preference for larger $f_{\rm max}$ values, as expected. It is interesting to note that in the fit over the full emulator range the inclusion of the CMASS data does not affect our $\gamma_f$ constraint, including not reducing the 1D marginalized uncertainty. However, there are several reasons why including additional data may not reduce 1D marginalized constraints. Firstly, the additional data may reduce the allowed parameter space in 14 dimensions without affecting the 1D constraints on a specific parameter. Additionally, the uncertainty in our measurement is limited by the emulator accuracy in several bins, notably the quadrupole and the large scale bins of the monopole and $w_p$, so a reduction of measurement uncertainty in these bins will not be reflected in the fit. Finally, the constraint on $\gamma_f$ seems to rely on the complimentary constraining of different scales and probes on parameter combinations such as $v_{\rm bc}$ and $\gamma_f$ (Fig.~\ref{fig:scales-probes}). The fit to CMASS+eBOSS has slightly less tension between the small and large scales than the eBOSS only measurement, so the overlap region remains the same size even though the uncertainty from separated scales has been reduced. This can be seen in the fit to the quasi-linear scales, where the combined CMASS+eBOSS sample gives a constraint of $f\sigma_8=0.384\pm0.036$. This constraint is consistent with the eBOSS only measurement from the quasi-linear scales, but because it is slightly lower it is in less tension with the fit over the full separation range.

\section{Discussion}\label{sec:discussion}

\subsection{Comparison to other measurements}\label{sec:disc_comparison}

\begin{figure*}
	\includegraphics[width=\textwidth]{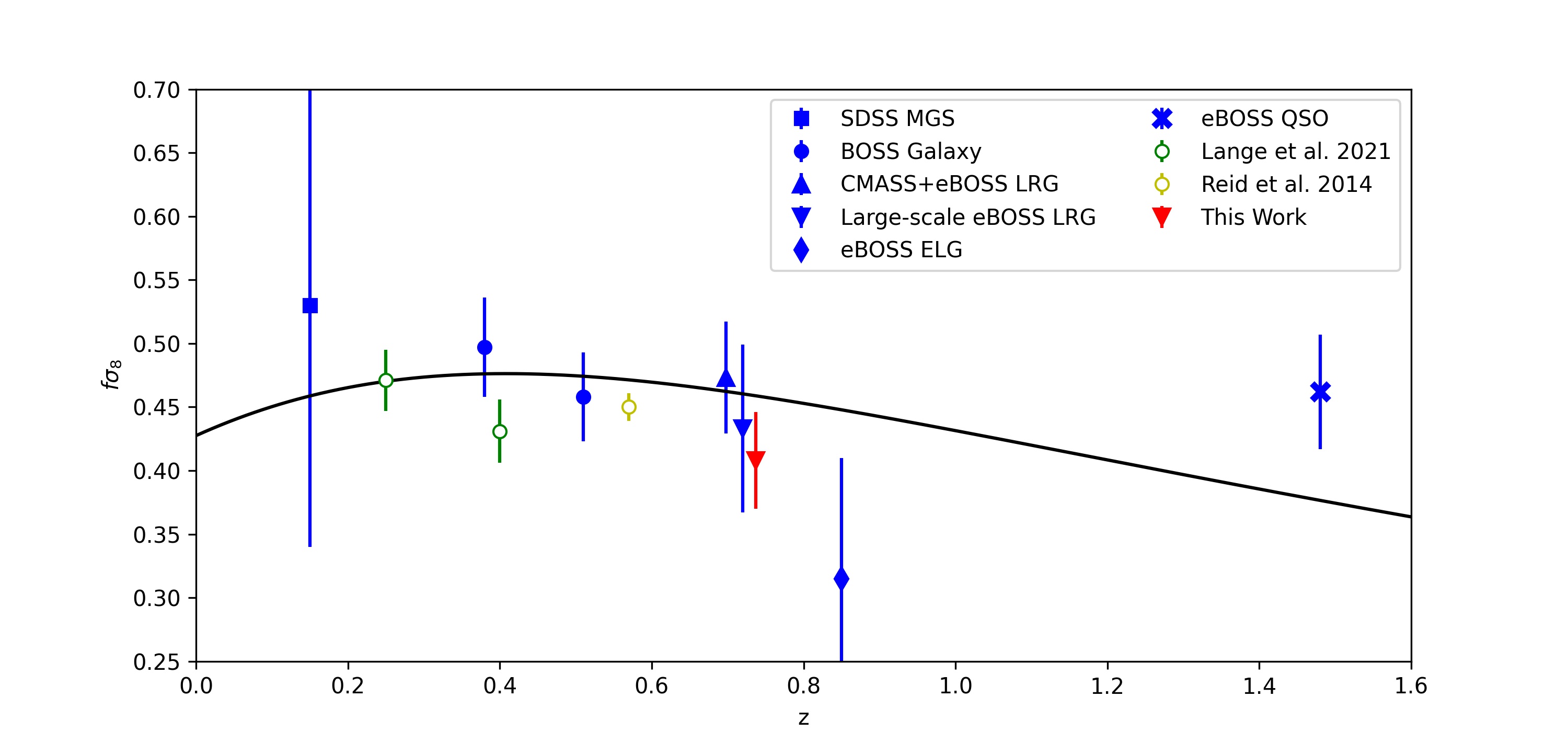}
    \caption{$f\sigma_8$ measurements from various SDSS samples. The blue points show the results of the standard large-scale analyses from the SDSS MGS \citep{Howlett_clustering_2015}, BOSS galaxies \citep{Alam:2017}, CMASS+eBOSS LRGs, eBOSS LRGs \citep{LRG_corr}, eBOSS ELGs \citep{demattia20a}, and eBOSS quasars \citep{neveux20a}. Our small-scale analysis of the eBOSS LRGs using only the quasi-linear regimes is shown in red. Empty coloured points show the results of small-scale analyses from the BOSS LOWZ sample (\citealt{Lange:2021}, green) and BOSS CMASS sample (\citealt{Reid:2014}, yellow) that included non-linear scales in the analysis. The black line shows the expected value of $f\sigma_8$ for a flat $\Lambda$CDM universe with best fit Planck2018 cosmology. The large-scale eBOSS LRG result is shifted in the x-axis to avoid overlap with the small-scale result from this work.}
    \label{fig:sdss-results}
\end{figure*}

We compare our result to other measurements of $f\sigma_8$ from galaxy clustering surveys in Fig.~\ref{fig:sdss-results}. Taken as a whole sample, there is clearly good consistency with the $\Lambda$CDM prediction. For the eBOSS LRGs, \citet{LRG_corr} analyzed pairs with separations between $25-130\mhmpc$, and obtained measurements of $f\sigma_8=0.446\pm0.066$ and $f\sigma_8=0.420\pm0.065$ depending on the RSD model used in the analysis (see Table B1 of \citealt{LRG_corr}). Our measurement is consistent with these results at around the ~$1-\sigma$ level, but has a factor of 1.7 improvement in the statistical error. Our measurement also continues the trend of galaxy clustering measurements of $f\sigma_8$ falling slightly below the prediction from observations of the CMB.

In Fig.~\ref{fig:sdss-results} we also compare our results to other attempts to measure $f\sigma_8$ on small scales. \citet{Reid:2014} used a similar parameterization as our analysis to measure $f\sigma_8$ from the small-scale clustering of the BOSS CMASS sample, and achieved the highest precision to date. However, due to the difficulty of modelling the non-linear regime \citet{Reid:2014} used a fixed cosmology, which has been shown by \citet{zhai_aemulus3} to significantly reduce the uncertainty. Conversely, \citet{Lange:2021} use a novel modelling method in their analysis of the BOSS LOWZ sample that does not require an emulator. It should also be noted that their model does not include an equivalent of our $\gamma_f$ parameter that allows the linear growth rate to change independently of the $\Lambda$CDM cosmology. Both of these analyses have split in linear and non-linear regimes differently than our analysis, which significantly affects the claimed uncertainty. By restricting our measurement of $f\sigma_8$ to only the quasi-linear scales our uncertainty increases by a factor of $\sim$1.5 compared to our fit over the full $0.1-60\mhmpc$ separation range, however we can be confident that what we are measuring is purely the linear growth rate, and so can be directly compared to other more standard large scale measurements. As shown in Sec.~\ref{sec:baseline_results} and Sec.~\ref{sec:scales_results}, using the full separation range significantly increases the tension with the result expect for $\Lambda$CDM, with the non-linear scales in greater disagreement with the expected value than the quasi-linear scales, however it is no longer clear if this tension arises from a discrepancy in the linear growth rate or a difference in the non-linear velocity field measured in the data using the emulator model.

It is interesting to note that \citet{Lange:2021} found a similar dependence on the measurement scales, with smaller scales preferring a smaller value of $f\sigma_8$. \citet{Lange:2021} also found that adding the projected correlation function to their fiducial measurement of the monopole, quadrupole, and hexadecapole reduced the best fit value of their lower redshift sample by $~1-\sigma$, but did not significantly affect the measurement from their higher redshift sample. Differences between the two analysis methods mean it is expected that there would be some variation in the impact of the different measurements and scales between our results. This is particularly true since \citet{Lange:2021} do not include a parameter comparable to our $\gamma_f$, given the importance of $w_p$ in breaking the $v_{\rm bc}-\gamma_f$ degeneracy in our analysis.

\subsection{Galaxy-halo connection parameters}

The parameter found to be most degenerate with our $\gamma_f$ constraint is $v_{\rm bc}$, the scaling of the velocity dispersion of centrals in the HOD framework (Fig.~\ref{fig:triangle-plot}). A lower value of $v_{\rm bc}$ corresponds to a larger $\gamma_f$, as expected in the non-linear regime since both parameters increase the observed velocity dispersion of galaxies (see Sec.~\ref{sec:zerr_test}). Our fit over the full $0.1-60 \mhmpc$ separation range strongly prefers a non-zero $v_{\rm bc}$ and low $\gamma_f$. However, our fit to the quasi-linear regime finds no discernible degeneracy between $v_{\rm bc}$ and $\gamma_f$ and recovers both a relatively large value of $\gamma_f$ and non-zero value of $v_{\rm bc}$, although the constraint on $v_{\rm bc}$ is weak to the small impact it has on those scales (Fig.~\ref{fig:scales-probes}). This result indicates that the degeneracy between $v_{\rm bc}$ and $f\sigma_8$ may illustrate the degree to which the non-linear scales affect the overall constraint. \citet{Lange:2021} also find a strong degeneracy between the velocity scaling of central galaxies and their constraint on $f\sigma_8$, with their higher redshift sample yielding $v_{\rm bc}>0$ and low $f\sigma_8$ compared to the $\Lambda$CDM prediction. \citet{Reid:2014} elected to fix the velocity of centrals to match that of the host halo, and find closer agreement with the $\Lambda$CDM expectation, which we also find when using a fixed $v_{\rm bc}=0$. $v_{\rm bc}>0$ indicates that a central galaxy is in motion relative to the centre of the host halo, either because the central galaxy is oscillating in the potential or because the system is not fully relaxed. Understanding the physical processes that would lead to this effect, especially if the process is redshift dependent, will be important for future analyses.

We also investigate the dependence of our measurement on the $f_{\rm max}$ parameter. Due to the strong degeneracy between $\sigma_{\log M}$ and $f_{\rm max}$, our fit to the data is broadly consistent with a wide range of values for $f_{\rm max}$ between 0.2 and 1, however there is a large peak at $f_{\rm max}=0.25$. A low value of $f_{\rm max}$ is not surprising for the eBOSS sample given the magnitude and color cuts made when selecting the target sample, particularly since the highest magnitude objects were removed. We do not find a degeneracy with $f\sigma_8$, so the lack of constraint on $\sigma_{\log M}$ and $f_{\rm max}$ is not expected to bias our measurement.

Numerical simulations have shown that the clustering of dark matter halos can depend on properties other than halo mass, a.k.a halo assembly bias (\citealt{Sheth_2004, Gao_2005, Harker_2006, Wechsler2006,Obuljen19}). This bias can propagate into the distribution of galaxies that live in these halos and thus introduce additional bias in the clustering measurement. In the analysis of BOSS galaxies over a wider redshift range \cite{Zhai_2021}, we enhance the basic HOD approach used here with an assembly bias model depending on the environment of dark matter halos. Although the results of that analysis imply the mild existence of assembly bias, there is a negligible impact on the cosmological constraint and measurement of structure growth rate. Therefore we exclude explicit modeling of assembly bias in this paper.

\begin{figure*}
    \centering
    \begin{subfigure}[b]{\columnwidth}
    	\includegraphics[width=\columnwidth]{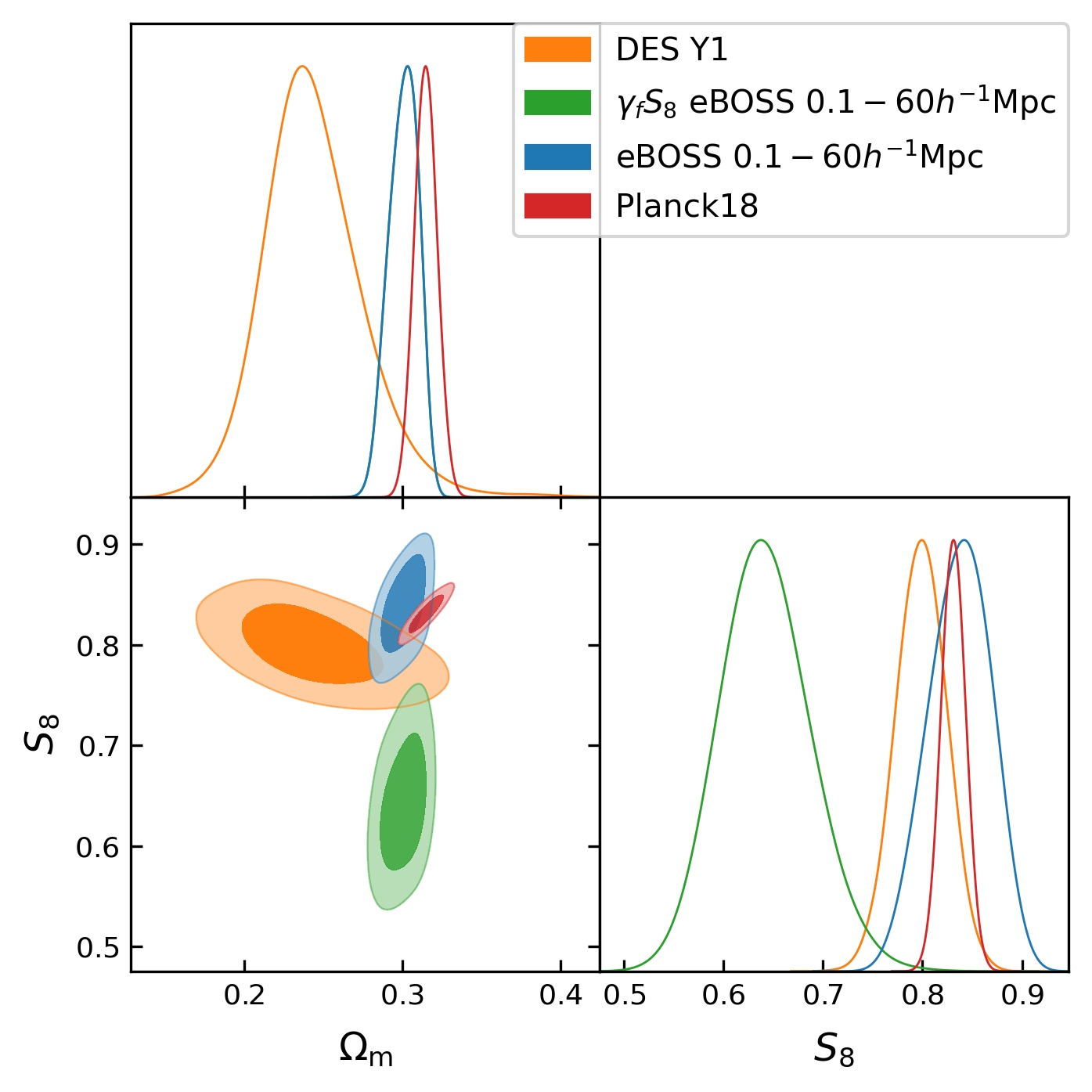}
        \label{fig:des-comp-full}
    \end{subfigure}
    \hfill
    \begin{subfigure}[b]{\columnwidth}
    	\includegraphics[width=\columnwidth]{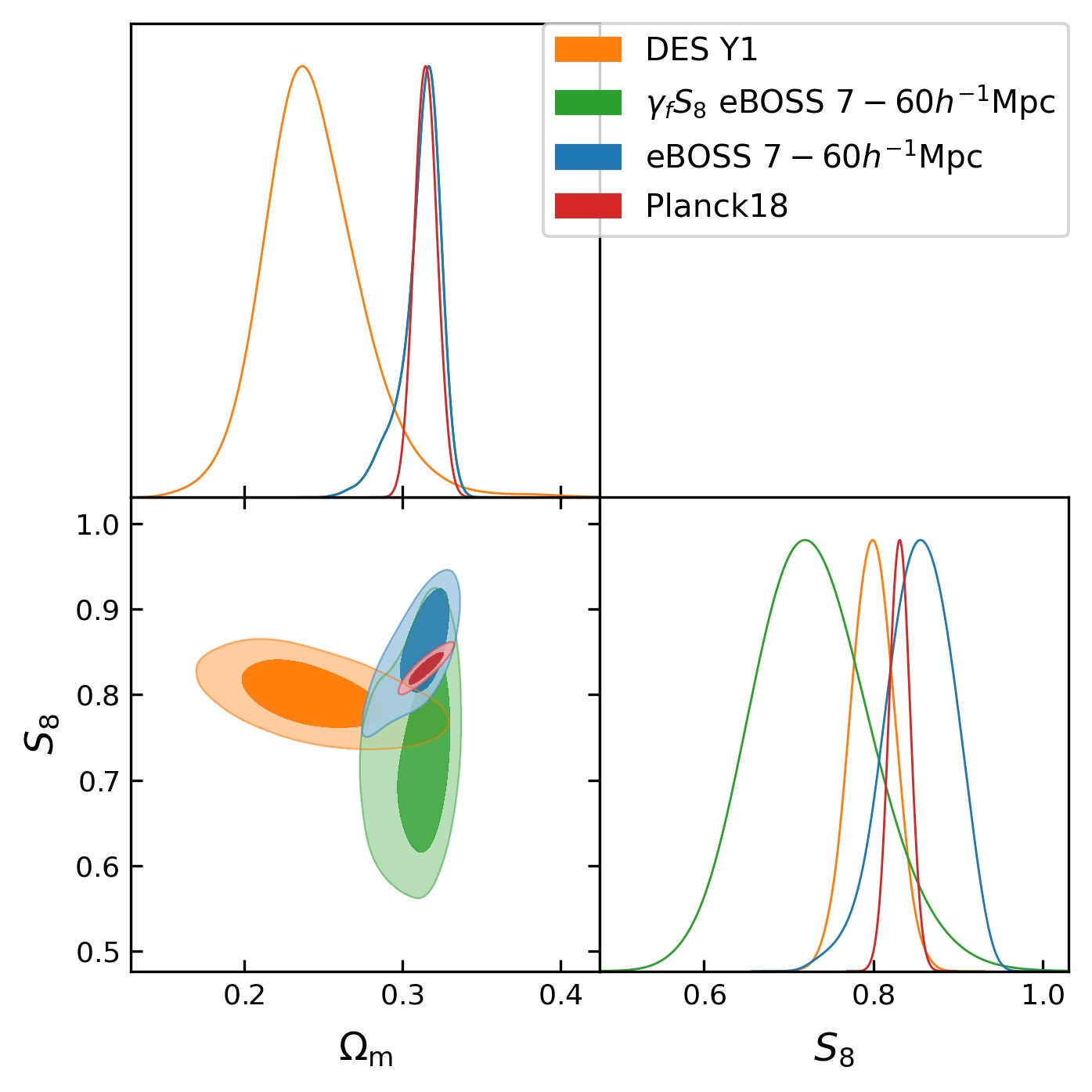}
        \label{fig:des-comp-lin}
    \end{subfigure}
        \caption{2D and 1D marginalized constraints on $\Omega_m$ and $S_8$ from our analysis (blue), the Dark Energy Survey (DES) year 1 results \citep{DES_Y1} (orange) and Planck 2018 results \citep{Aghanim:2019ame, Aghanim:2018oex} (red). Since our low value of $f\sigma_8$ mostly comes from $\gamma_f$, we also plot $\gamma_f S_8$ (green) for our fit, which shows the constraint we would have if the low value of $f\sigma_8$ came entirely from the $\sigma_8$ value. \textit{Left:} Results of our full fit to all scales. \textit{Right:} Results from only the quasi-linear scales used to constrain the linear growth rate.}
        \label{fig:des-comp}
\end{figure*}

\subsection{Comparison to tension from lensing surveys}

It is interesting to note that we obtain a lower value of $f\sigma_8$ than expected from Planck measurements, given the current $\sigma_8$-tension between Planck and weak lensing surveys and the low amplitude of the galaxy-galaxy lensing amplitude measured using the BOSS CMASS sample by \citet{Leauthaud:2017}, since both tensions could be resolved by a lower value of $\sigma_8$ than that measured by Planck. To see approximately how our result might relate to this tension we compare the constraints on $S_8=\sigma_8(\Omega_M/0.3)^{0.5}$ for the DES Y1 results \citep{DES_Y1}, Planck 2018 \citep{Planck-2018-params}, and our results (Fig.~\ref{fig:des-comp}). The left panel shows our measurement using the full separation range, while the right panel shows our measurement from the quasi-linear scales only. Our constraint, shown as the blue contour, is consistent with both the DES Y1 and Planck results in both cases. However, it is important to note that our low value of $f\sigma_8$ comes almost entirely from $\gamma_f<1$, which reduces the magnitude of peculiar velocities in the simulation without affecting the amplitude of fluctuations, $\sigma_8$. If the low value of $f\sigma_8$ we measure was due to the value of $\sigma_8$ instead then the constraint would shift down the $S_8$ axis, shown as a green contour. For our measurement from the quasi-linear scales this shift maintains consistency with both DES Y1 and Planck 2018, however for our fit to all scales this shift puts the green constraint in tension with the Planck results, and in more mild disagreement with the DES results. This result may indicate that the increased tension we find from the non-linear scales may be caused by an issue with the HOD model, rather than a purely cosmological tension.

\subsection{Emulator robustness and potential improvements}

We have performed rigorous tests of the emulator performance (see Sec.~\ref{sec:model_tests}), and found that the model performs well when fit to an independent simulation and galaxy-halo connection prescription. We also find that a model that assumes all central galaxies are observed leads to a systematic bias in the recovered cosmological parameters if the actual fractional occupation of centrals is lower than 0.6. We correct this bias by adding the parameter $f_{\rm max}$ to the emulator, and verify that the full emulator gives an unbiased measurement for $0.2<f_{\rm max}<1.0$. We also identify the redshift uncertainty as a source of systematic bias on non-linear scales, with a redshift uncertainty missing from the model leading to an offset in $\gamma_f$ to larger values by more than half of the statistical error for the eBOSS sample. This is a significant concern for future small-scale analyses, and will require careful attention due to the difficulties in implementing a redshift dependent effect in a model constructed at a single redshift. The redshift uncertainty has also been found to scale with redshift, so it will be an even greater concern for future large surveys at high redshift such as DESI \citep{desi1,desi2} and Euclid \citep{Laureijs-Euclid}.

Our measurement of the clustering within the eBOSS LRG sample also meets or exceeds the emulator precision in several of the measurement bins (see Fig.~\ref{fig:error_components}), showing the importance of improving the model precision for future surveys. This must be balanced against ensuring there are sufficient bins to yield a well defined fit, given the number of model parameters (see Sec. ~\ref{sec:results_N_params}). Finally, careful attention must be given to the non-linear scales, and identifying what information can be used to constrain the linear growth rate. A key aspect includes ensuring the performance of the HOD model on these scales, and investigating the effect of baryonic physics.

\section{Summary} \label{sec:summary}

We have measured the growth rate of structure from the small-scale clustering of the eBOSS LRG sample corrected by PIP weights and modelled using the \textsc{aemulus} cosmological emulator. Using the quasi-linear scales of our measurement range we obtain a measurement of $f\sigma_8(z=0.737)=0.408\pm0.038$, which is $1.4\sigma$ lower than the value expected from 2018 Planck data for a flat $\Lambda$CDM model. Our measurement is a significant improvement over more standard measurements made using only the large scale modes, achieving a level of precision that is 1.7 times better than the large-scale analysis of the same sample. Using the full separation range of our measurement we find a $4.5\sigma$ tension in the amplitude of the halo velocity field with the expectation for a $\Lambda$CDM universe. This tension is driven by the non-linear scales of our analysis and so may not be well modelled by a change in the linear growth rate, but may instead reflect a breakdown in the HOD model used in the emulator.

We perform a robust check of possible sources of systematic error not included in previous analyses. We find that using a model that assumes all central galaxies are observed leads to a systematic bias if the actual occupation of centrals is lower; a fractional occupation of $f_{\rm max}\leq0.6$. We also investigate the effect of redshift uncertainty, and find that the presence of a velocity shift from redshift uncertainty in the data that is not included in the model results in a higher measurement of $\gamma_f$ with an offset of $\sim0.5\sigma$, where $\sigma$ is the typical statistical error. This effect is caused by the degeneracy between the increased velocity dispersion due to the redshift uncertainty and the random motions of the halos in the non-linear regime. Lastly, we investigate the consistency between the non-linear and quasi-linear scales of our analysis. While we find them to be consistent within the statistical error, there is a trend to lower $\gamma_f$ on non-linear scales, which increases the tension with the expectation from $\Lambda$CDM for the fit to all scales. This result highlights the importance of distinguishing between results obtained from the linear scales and thus directly constraining the linear growth rate $f\sigma_8$, and those that include non-linear scales and may have a non-linear dependence on the linear growth rate together with a dependence on other factors.

While our results are consistent with the expectation from Planck 2018 parameter constraints, we are also consistent with recent weak lensing results giving a low value of $S_8$, particularly if our low value of $f\sigma_8$ was driven by an adjustment to $\sigma_8$. In light of these lensing results and the mild disagreement we find with Planck expectations, extending this type of analysis to future surveys including DESI and Euclid will be an important area of future research. With considerably larger samples and probing a different redshift range, the improvement in precision from moving to smaller scales will be key to achieving optimal constraints and identifying or rejecting a tension in the growth rate of cosmic structure.

\section*{Acknowledgements}

ZZ is supported in part by NASA grant 15-WFIRST15-0008, Cosmology with the High Latitude Survey Roman Science Investigation Team (SIT). JLT acknowledges the support of NSF AAG grant 2009291. MJC, FGM, ZZ \& WJP acknowledge financial support from the Canadian Space Agency (CSA) and the Natural Sciences and Engineering Research Council of Canada (NSERC). G.R. acknowledges support from the National Research Foundation of Korea (NRF) through Grants No. 2017R1E1A1A01077508 and  No.2020R1A2C1005655 funded by the Korean Ministry of Education, Science and Technology (MoEST).

Research at Perimeter Institute is supported in part by the Government of Canada through the Department of Innovation, Science and Economic Development Canada and by the Province of Ontario through the Ministry of Colleges and Universities.

This research was enabled in part by support provided by Compute Ontario (www.computeontario.ca) and Compute Canada (www.computecanada.ca).

Funding for the Sloan Digital Sky Survey IV has been provided by the Alfred P. Sloan Foundation, the U.S. Department of Energy Office of Science, and the Participating Institutions. SDSS-IV acknowledges support and resources from the Center for High 
Performance Computing at the University of Utah. The SDSS 
website is www.sdss.org. SDSS-IV is managed by the Astrophysical Research Consortium for the Participating Institutions 
of the SDSS Collaboration including 
the Brazilian Participation Group, 
the Carnegie Institution for Science, 
Carnegie Mellon University, Center for 
Astrophysics | Harvard \& 
Smithsonian, the Chilean Participation 
Group, the French Participation Group, 
Instituto de Astrof\'isica de 
Canarias, The Johns Hopkins 
University, Kavli Institute for the 
Physics and Mathematics of the 
Universe (IPMU) / University of 
Tokyo, the Korean Participation Group, 
Lawrence Berkeley National Laboratory, 
Leibniz Institut f\"ur Astrophysik 
Potsdam (AIP),  Max-Planck-Institut 
f\"ur Astronomie (MPIA Heidelberg), 
Max-Planck-Institut f\"ur 
Astrophysik (MPA Garching), 
Max-Planck-Institut f\"ur 
Extraterrestrische Physik (MPE), 
National Astronomical Observatories of 
China, New Mexico State University, 
New York University, University of 
Notre Dame, Observat\'ario 
Nacional / MCTI, The Ohio State 
University, Pennsylvania State 
University, Shanghai 
Astronomical Observatory, United 
Kingdom Participation Group, 
Universidad Nacional Aut\'onoma 
de M\'exico, University of Arizona, 
University of Colorado Boulder, 
University of Oxford, University of 
Portsmouth, University of Utah, 
University of Virginia, University 
of Washington, University of 
Wisconsin, Vanderbilt University, 
and Yale University.

\section*{Data Availability}

The eBOSS galaxy and random catalogues are publicly available at: https://data.sdss.org/sas/dr16/eboss/lss/catalogs/DR16/ with a description here: https://www.sdss.org/dr16/spectro/lss/ We used the \textsc{aemulus} emulator, which is available here: https://aemulusproject.github.io, and the Cobaya package, which is available here: https://github.com/CobayaSampler


\typeout{}
\bibliographystyle{mnras}
\bibliography{eboss_small_scale_rsd}

\begin{thebibliography}{}
\makeatletter
\relax
\def\mn@urlcharsother{\let\do\@makeother \do\$\do\&\do\#\do\^\do\_\do\%\do\~}
\def\mn@doi{\begingroup\mn@urlcharsother \@ifnextchar [ {\mn@doi@}
  {\mn@doi@[]}}
\def\mn@doi@[#1]#2{\def\@tempa{#1}\ifx\@tempa\@empty \href
  {http://dx.doi.org/#2} {doi:#2}\else \href {http://dx.doi.org/#2} {#1}\fi
  \endgroup}
\def\mn@eprint#1#2{\mn@eprint@#1:#2::\@nil}
\def\mn@eprint@arXiv#1{\href {http://arxiv.org/abs/#1} {{\tt arXiv:#1}}}
\def\mn@eprint@dblp#1{\href {http://dblp.uni-trier.de/rec/bibtex/#1.xml}
  {dblp:#1}}
\def\mn@eprint@#1:#2:#3:#4\@nil{\def\@tempa {#1}\def\@tempb {#2}\def\@tempc
  {#3}\ifx \@tempc \@empty \let \@tempc \@tempb \let \@tempb \@tempa \fi \ifx
  \@tempb \@empty \def\@tempb {arXiv}\fi \@ifundefined
  {mn@eprint@\@tempb}{\@tempb:\@tempc}{\expandafter \expandafter \csname
  mn@eprint@\@tempb\endcsname \expandafter{\@tempc}}}

\bibitem[\protect\citeauthoryear{{Abbott} et~al.,}{{Abbott}
  et~al.}{2018}]{DES_Y1}
{Abbott} T.~M.~C.,  et~al., 2018, \mn@doi [\prd] {10.1103/PhysRevD.98.043526},
  \href {https://ui.adsabs.harvard.edu/abs/2018PhRvD..98d3526A} {98, 043526}

\bibitem[\protect\citeauthoryear{{Ahumada} et~al.,}{{Ahumada}
  et~al.}{2020}]{DR16}
{Ahumada} R.,  et~al., 2020, \mn@doi [\apjs] {10.3847/1538-4365/ab929e}, \href
  {https://ui.adsabs.harvard.edu/abs/2020ApJS..249....3A} {249, 3}

\bibitem[\protect\citeauthoryear{{Alam} et~al.,}{{Alam}
  et~al.}{2017}]{Alam:2017}
{Alam} S.,  et~al., 2017, \mn@doi [\mnras] {10.1093/mnras/stx721}, \href
  {https://ui.adsabs.harvard.edu/abs/2017MNRAS.470.2617A} {470, 2617}

\bibitem[\protect\citeauthoryear{{Alam} et~al.,}{{Alam}
  et~al.}{2021}]{eBOSS_Cosmology}
{Alam} S.,  et~al., 2021, \mn@doi [\prd] {10.1103/PhysRevD.103.083533}, \href
  {https://ui.adsabs.harvard.edu/abs/2021PhRvD.103h3533A} {103, 083533}

\bibitem[\protect\citeauthoryear{{Albareti} et~al.,}{{Albareti}
  et~al.}{2017}]{Albareti:2017}
{Albareti} F.~D.,  et~al., 2017, \mn@doi [\apjs] {10.3847/1538-4365/aa8992},
  \href {https://ui.adsabs.harvard.edu/abs/2017ApJS..233...25A} {233, 25}

\bibitem[\protect\citeauthoryear{Alcock \& Paczynski}{Alcock \&
  Paczynski}{1979}]{AP79}
Alcock C.,  Paczynski B.,  1979, \mn@doi [Nature] {10.1038/281358a0}, 281, 358

\bibitem[\protect\citeauthoryear{{Bautista} et~al.,}{{Bautista}
  et~al.}{2021}]{LRG_corr}
{Bautista} J.~E.,  et~al., 2021, \mn@doi [\mnras] {10.1093/mnras/staa2800},
  \href {https://ui.adsabs.harvard.edu/abs/2021MNRAS.500..736B} {500, 736}

\bibitem[\protect\citeauthoryear{Beutler et~al.,}{Beutler
  et~al.}{2012}]{beutler_6df_2012}
Beutler F.,  et~al., 2012, \mn@doi [Monthly Notices of the Royal Astronomical
  Society] {10.1111/j.1365-2966.2012.21136.x}, 423, 3430

\bibitem[\protect\citeauthoryear{Beutler et~al.,}{Beutler
  et~al.}{2017}]{beutler_clustering_2017}
Beutler F.,  et~al., 2017, \mn@doi [Monthly Notices of the Royal Astronomical
  Society] {10.1093/mnras/stw3298}, 466, 2242

\bibitem[\protect\citeauthoryear{{Bianchi} \& {Percival}}{{Bianchi} \&
  {Percival}}{2017}]{Bianchi:2017}
{Bianchi} D.,  {Percival} W.~J.,  2017, \mn@doi [\mnras]
  {10.1093/mnras/stx2053}, \href
  {https://ui.adsabs.harvard.edu/abs/2017MNRAS.472.1106B} {472, 1106}

\bibitem[\protect\citeauthoryear{Blake et~al.,}{Blake
  et~al.}{2011}]{blake_wigglez_2011}
Blake C.,  et~al., 2011, \mn@doi [Monthly Notices of the Royal Astronomical
  Society] {10.1111/j.1365-2966.2011.18903.x}, 415, 2876

\bibitem[\protect\citeauthoryear{Blanton et~al.,}{Blanton
  et~al.}{2017}]{blanton_sloan_2017}
Blanton M.~R.,  et~al., 2017, \mn@doi [The Astronomical Journal]
  {10.3847/1538-3881/aa7567}, 154, 28

\bibitem[\protect\citeauthoryear{{Bolton} et~al.,}{{Bolton}
  et~al.}{2012}]{Bolton:2012}
{Bolton} A.~S.,  et~al., 2012, \mn@doi [\aj] {10.1088/0004-6256/144/5/144},
  \href {https://ui.adsabs.harvard.edu/abs/2012AJ....144..144B} {144, 144}

\bibitem[\protect\citeauthoryear{{DESI Collaboration} et~al.,}{{DESI
  Collaboration} et~al.}{2016a}]{desi2}
{DESI Collaboration} et~al., 2016a, arXiv e-prints, \href
  {https://ui.adsabs.harvard.edu/abs/2016arXiv161100036D} {p. arXiv:1611.00036}

\bibitem[\protect\citeauthoryear{{DESI Collaboration} et~al.,}{{DESI
  Collaboration} et~al.}{2016b}]{desi1}
{DESI Collaboration} et~al., 2016b, arXiv e-prints, \href
  {https://ui.adsabs.harvard.edu/abs/2016arXiv161100037D} {p. arXiv:1611.00037}

\bibitem[\protect\citeauthoryear{Dawson et~al.,}{Dawson
  et~al.}{2013}]{dawson_baryon_2013}
Dawson K.~S.,  et~al., 2013, \mn@doi [The Astronomical Journal]
  {10.1088/0004-6256/145/1/10}, 145, 10

\bibitem[\protect\citeauthoryear{Dawson et~al.,}{Dawson
  et~al.}{2016}]{dawson_sdss-iv_2016}
Dawson K.~S.,  et~al., 2016, \mn@doi [The Astronomical Journal]
  {10.3847/0004-6256/151/2/44}, 151, 44

\bibitem[\protect\citeauthoryear{{DeRose} et~al.,}{{DeRose}
  et~al.}{2019}]{derose_aemulus1}
{DeRose} J.,  et~al., 2019, \mn@doi [\apj] {10.3847/1538-4357/ab1085}, \href
  {https://ui.adsabs.harvard.edu/abs/2019ApJ...875...69D} {875, 69}

\bibitem[\protect\citeauthoryear{{Eifler}, {Kilbinger}  \&
  {Schneider}}{{Eifler} et~al.}{2008}]{Eifler:2008}
{Eifler} T.,  {Kilbinger} M.,   {Schneider} P.,  2008, \mn@doi [\aap]
  {10.1051/0004-6361:20078573}, \href
  {https://ui.adsabs.harvard.edu/abs/2008A&A...482....9E} {482, 9}

\bibitem[\protect\citeauthoryear{Eisenstein et~al.,}{Eisenstein
  et~al.}{2011}]{eisenstein_sdss-iii:_2011}
Eisenstein D.~J.,  et~al., 2011, \mn@doi [The Astronomical Journal]
  {10.1088/0004-6256/142/3/72;}, 142, 72

\bibitem[\protect\citeauthoryear{{Feldman}, {Kaiser}  \& {Peacock}}{{Feldman}
  et~al.}{1994}]{Feldman:1994}
{Feldman} H.~A.,  {Kaiser} N.,   {Peacock} J.~A.,  1994, \mn@doi [\apj]
  {10.1086/174036}, \href {http://adsabs.harvard.edu/abs/1994ApJ...426...23F}
  {426, 23}

\bibitem[\protect\citeauthoryear{Ferreira}{Ferreira}{2019}]{ferreira_cosmological_2019}
Ferreira P.~G.,  2019, \mn@doi [Annual Review of Astronomy and Astrophysics]
  {10.1146/annurev-astro-091918-104423}, 57, 335

\bibitem[\protect\citeauthoryear{{Gao}, {Springel}  \& {White}}{{Gao}
  et~al.}{2005}]{Gao_2005}
{Gao} L.,  {Springel} V.,   {White} S.~D.~M.,  2005, \mn@doi [\mnras]
  {10.1111/j.1745-3933.2005.00084.x}, \href
  {http://adsabs.harvard.edu/abs/2005MNRAS.363L..66G} {363, L66}

\bibitem[\protect\citeauthoryear{{Gil-Mar{\'\i}n} et~al.,}{{Gil-Mar{\'\i}n}
  et~al.}{2020}]{gil-marin20a}
{Gil-Mar{\'\i}n} H.,  et~al., 2020, \mn@doi [\mnras] {10.1093/mnras/staa2455},
  \href {https://ui.adsabs.harvard.edu/abs/2020MNRAS.498.2492G} {498, 2492}

\bibitem[\protect\citeauthoryear{Grieb et~al.,}{Grieb
  et~al.}{2017}]{grieb_clustering_2017}
Grieb J.~N.,  et~al., 2017, \mn@doi [Monthly Notices of the Royal Astronomical
  Society] {10.1093/mnras/stw3384}, 467, 2085

\bibitem[\protect\citeauthoryear{{Gunn} et~al.,}{{Gunn}
  et~al.}{2006}]{Gunn:2006}
{Gunn} J.~E.,  et~al., 2006, \mn@doi [\aj] {10.1086/500975}, \href
  {https://ui.adsabs.harvard.edu/abs/2006AJ....131.2332G} {131, 2332}

\bibitem[\protect\citeauthoryear{Guzzo et~al.,}{Guzzo
  et~al.}{2008}]{guzzo_test_2008}
Guzzo L.,  et~al., 2008, \mn@doi [Nature] {10.1038/nature06555}, 451, 541

\bibitem[\protect\citeauthoryear{{Harker}, {Cole}, {Helly}, {Frenk}  \&
  {Jenkins}}{{Harker} et~al.}{2006}]{Harker_2006}
{Harker} G.,  {Cole} S.,  {Helly} J.,  {Frenk} C.,   {Jenkins} A.,  2006,
  \mn@doi [\mnras] {10.1111/j.1365-2966.2006.10022.x}, \href
  {http://adsabs.harvard.edu/abs/2006MNRAS.367.1039H} {367, 1039}

\bibitem[\protect\citeauthoryear{{Hartlap}, {Simon}  \& {Schneider}}{{Hartlap}
  et~al.}{2007}]{Hartlap:2007}
{Hartlap} J.,  {Simon} P.,   {Schneider} P.,  2007, \mn@doi [\aap]
  {10.1051/0004-6361:20066170}, \href
  {https://ui.adsabs.harvard.edu/abs/2007A&A...464..399H} {464, 399}

\bibitem[\protect\citeauthoryear{{Hearin} et~al.,}{{Hearin}
  et~al.}{2017}]{halotools}
{Hearin} A.~P.,  et~al., 2017, \mn@doi [\aj] {10.3847/1538-3881/aa859f}, \href
  {https://ui.adsabs.harvard.edu/abs/2017AJ....154..190H} {154, 190}

\bibitem[\protect\citeauthoryear{{Hinshaw} et~al.,}{{Hinshaw}
  et~al.}{2013}]{WMAP-final-params}
{Hinshaw} G.,  et~al., 2013, \mn@doi [\apjs] {10.1088/0067-0049/208/2/19},
  \href {https://ui.adsabs.harvard.edu/abs/2013ApJS..208...19H} {208, 19}

\bibitem[\protect\citeauthoryear{{Hou} et~al.,}{{Hou} et~al.}{2021}]{hou20a}
{Hou} J.,  et~al., 2021, \mn@doi [\mnras] {10.1093/mnras/staa3234}, \href
  {https://ui.adsabs.harvard.edu/abs/2021MNRAS.500.1201H} {500, 1201}

\bibitem[\protect\citeauthoryear{Howlett, Lewis, Hall  \& Challinor}{Howlett
  et~al.}{2012}]{Howlett:2012mh}
Howlett C.,  Lewis A.,  Hall A.,   Challinor A.,  2012, \mn@doi [JCAP]
  {10.1088/1475-7516/2012/04/027}, 1204, 027

\bibitem[\protect\citeauthoryear{Howlett, Ross, Samushia, Percival  \&
  Manera}{Howlett et~al.}{2015}]{Howlett_clustering_2015}
Howlett C.,  Ross A.~J.,  Samushia L.,  Percival W.~J.,   Manera M.,  2015,
  \mn@doi [Monthly Notices of the Royal Astronomical Society]
  {10.1093/mnras/stu2693}, 449, 848

\bibitem[\protect\citeauthoryear{{Kaiser}}{{Kaiser}}{1987}]{Kaiser:1987}
{Kaiser} N.,  1987, \mn@doi [\mnras] {10.1093/mnras/227.1.1}, \href
  {http://adsabs.harvard.edu/abs/1987MNRAS.227....1K} {227, 1}

\bibitem[\protect\citeauthoryear{{Krause}, {Hirata}, {Martin}, {Neill}  \&
  {Wyder}}{{Krause} et~al.}{2013}]{Krauss:2013}
{Krause} E.,  {Hirata} C.~M.,  {Martin} C.,  {Neill} J.~D.,   {Wyder} T.~K.,
  2013, \mn@doi [\mnras] {10.1093/mnras/sts221}, \href
  {https://ui.adsabs.harvard.edu/abs/2013MNRAS.428.2548K} {428, 2548}

\bibitem[\protect\citeauthoryear{{Landy} \& {Szalay}}{{Landy} \&
  {Szalay}}{1993}]{landy93}
{Landy} S.~D.,  {Szalay} A.~S.,  1993, \mn@doi [\apj] {10.1086/172900}, \href
  {https://ui.adsabs.harvard.edu/abs/1993ApJ...412...64L} {412, 64}

\bibitem[\protect\citeauthoryear{{Lang}, {Hogg}  \& {Schlegel}}{{Lang}
  et~al.}{2016}]{Lang:2016}
{Lang} D.,  {Hogg} D.~W.,   {Schlegel} D.~J.,  2016, \mn@doi [\aj]
  {10.3847/0004-6256/151/2/36}, \href
  {https://ui.adsabs.harvard.edu/abs/2016AJ....151...36L} {151, 36}

\bibitem[\protect\citeauthoryear{{Lange}, {Hearin}, {Leauthaud}, {van den
  Bosch}, {Guo}  \& {DeRose}}{{Lange} et~al.}{2022}]{Lange:2021}
{Lange} J.~U.,  {Hearin} A.~P.,  {Leauthaud} A.,  {van den Bosch} F.~C.,  {Guo}
  H.,   {DeRose} J.,  2022, \mn@doi [\mnras] {10.1093/mnras/stab3111}, \href
  {https://ui.adsabs.harvard.edu/abs/2022MNRAS.509.1779L} {509, 1779}

\bibitem[\protect\citeauthoryear{{Laureijs} et~al.,}{{Laureijs}
  et~al.}{2011}]{Laureijs-Euclid}
{Laureijs} R.,  et~al., 2011, arXiv e-prints, \href
  {https://ui.adsabs.harvard.edu/abs/2011arXiv1110.3193L} {p. arXiv:1110.3193}

\bibitem[\protect\citeauthoryear{{Leauthaud} et~al.,}{{Leauthaud}
  et~al.}{2017}]{Leauthaud:2017}
{Leauthaud} A.,  et~al., 2017, \mn@doi [\mnras] {10.1093/mnras/stx258}, \href
  {https://ui.adsabs.harvard.edu/abs/2017MNRAS.467.3024L} {467, 3024}

\bibitem[\protect\citeauthoryear{Lewis}{Lewis}{2013}]{Lewis:2013hha}
Lewis A.,  2013, \mn@doi [Phys. Rev.] {10.1103/PhysRevD.87.103529}, D87, 103529

\bibitem[\protect\citeauthoryear{Lewis \& Bridle}{Lewis \&
  Bridle}{2002}]{Lewis:2002ah}
Lewis A.,  Bridle S.,  2002, \mn@doi [Phys. Rev.] {10.1103/PhysRevD.66.103511},
  D66, 103511

\bibitem[\protect\citeauthoryear{Lewis, Challinor  \& Lasenby}{Lewis
  et~al.}{2000}]{Lewis:1999bs}
Lewis A.,  Challinor A.,   Lasenby A.,  2000, \mn@doi [Astrophys. J.]
  {10.1086/309179}, 538, 473

\bibitem[\protect\citeauthoryear{{Lyke} et~al.,}{{Lyke} et~al.}{2020}]{lyke20a}
{Lyke} B.~W.,  et~al., 2020, \mn@doi [\apjs] {10.3847/1538-4365/aba623}, \href
  {https://ui.adsabs.harvard.edu/abs/2020ApJS..250....8L} {250, 8}

\bibitem[\protect\citeauthoryear{{Mohammad}, {de la Torre}, {Bianchi}, {Guzzo}
  \& {Peacock}}{{Mohammad} et~al.}{2016}]{mohammad16}
{Mohammad} F.~G.,  {de la Torre} S.,  {Bianchi} D.,  {Guzzo} L.,   {Peacock}
  J.~A.,  2016, \mn@doi [\mnras] {10.1093/mnras/stw411}, \href
  {https://ui.adsabs.harvard.edu/abs/2016MNRAS.458.1948M} {458, 1948}

\bibitem[\protect\citeauthoryear{{Mohammad} et~al.,}{{Mohammad}
  et~al.}{2020}]{Mohammad:2020}
{Mohammad} F.~G.,  et~al., 2020, \mn@doi [\mnras] {10.1093/mnras/staa2344},
  \href {https://ui.adsabs.harvard.edu/abs/2020MNRAS.498..128M} {498, 128}

\bibitem[\protect\citeauthoryear{{Navarro}, {Frenk}  \& {White}}{{Navarro}
  et~al.}{1996}]{NFW}
{Navarro} J.~F.,  {Frenk} C.~S.,   {White} S. D.~M.,  1996, \mn@doi [\apj]
  {10.1086/177173}, \href
  {https://ui.adsabs.harvard.edu/abs/1996ApJ...462..563N} {462, 563}

\bibitem[\protect\citeauthoryear{{Neal}}{{Neal}}{2005}]{Neal:2005}
{Neal} R.~M.,  2005, ArXiv Mathematics e-prints

\bibitem[\protect\citeauthoryear{{Neveux} et~al.,}{{Neveux}
  et~al.}{2020}]{neveux20a}
{Neveux} R.,  et~al., 2020, \mn@doi [\mnras] {10.1093/mnras/staa2780}, \href
  {https://ui.adsabs.harvard.edu/abs/2020MNRAS.499..210N} {499, 210}

\bibitem[\protect\citeauthoryear{{Obuljen}, {Dalal}  \& {Percival}}{{Obuljen}
  et~al.}{2019}]{Obuljen19}
{Obuljen} A.,  {Dalal} N.,   {Percival} W.~J.,  2019, \mn@doi [\jcap]
  {10.1088/1475-7516/2019/10/020}, \href
  {https://ui.adsabs.harvard.edu/abs/2019JCAP...10..020O} {2019, 020}

\bibitem[\protect\citeauthoryear{Okumura et~al.,}{Okumura
  et~al.}{2016}]{okumura_subaru_2016}
Okumura T.,  et~al., 2016, \mn@doi [Publications of the Astronomical Society of
  Japan] {10.1093/pasj/psw029}, 68, 38

\bibitem[\protect\citeauthoryear{{Percival} \& {Bianchi}}{{Percival} \&
  {Bianchi}}{2017}]{Percival:2017}
{Percival} W.~J.,  {Bianchi} D.,  2017, \mn@doi [\mnras]
  {10.1093/mnrasl/slx135}, \href
  {https://ui.adsabs.harvard.edu/abs/2017MNRAS.472L..40P} {472, L40}

\bibitem[\protect\citeauthoryear{Pezzotta et~al.,}{Pezzotta
  et~al.}{2017}]{pezzotta_vimos_2017}
Pezzotta A.,  et~al., 2017, \mn@doi [Astronomy and Astrophysics]
  {10.1051/0004-6361/201630295}, 604, A33

\bibitem[\protect\citeauthoryear{{Planck Collaboration} et~al.,}{{Planck
  Collaboration} et~al.}{2020a}]{Aghanim:2019ame}
{Planck Collaboration} et~al., 2020a, \mn@doi [\aap]
  {10.1051/0004-6361/201936386}, \href
  {https://ui.adsabs.harvard.edu/abs/2020A&A...641A...5P} {641, A5}

\bibitem[\protect\citeauthoryear{{Planck Collaboration} et~al.,}{{Planck
  Collaboration} et~al.}{2020b}]{Planck-2018-params}
{Planck Collaboration} et~al., 2020b, \mn@doi [\aap]
  {10.1051/0004-6361/201833910}, \href
  {https://ui.adsabs.harvard.edu/abs/2020A&A...641A...6P} {641, A6}

\bibitem[\protect\citeauthoryear{{Planck Collaboration} et~al.,}{{Planck
  Collaboration} et~al.}{2020c}]{Aghanim:2018oex}
{Planck Collaboration} et~al., 2020c, \mn@doi [\aap]
  {10.1051/0004-6361/201833886}, \href
  {https://ui.adsabs.harvard.edu/abs/2020A&A...641A...8P} {641, A8}

\bibitem[\protect\citeauthoryear{{Prakash} et~al.,}{{Prakash}
  et~al.}{2016}]{Prakash:2016}
{Prakash} A.,  et~al., 2016, \mn@doi [\apjs] {10.3847/0067-0049/224/2/34},
  \href {https://ui.adsabs.harvard.edu/abs/2016ApJS..224...34P} {224, 34}

\bibitem[\protect\citeauthoryear{{Raichoor} et~al.,}{{Raichoor}
  et~al.}{2021}]{raichoor20a}
{Raichoor} A.,  et~al., 2021, \mn@doi [\mnras] {10.1093/mnras/staa3336}, \href
  {https://ui.adsabs.harvard.edu/abs/2021MNRAS.500.3254R} {500, 3254}

\bibitem[\protect\citeauthoryear{{Reid}, {Seo}, {Leauthaud}, {Tinker}  \&
  {White}}{{Reid} et~al.}{2014}]{Reid:2014}
{Reid} B.~A.,  {Seo} H.-J.,  {Leauthaud} A.,  {Tinker} J.~L.,   {White} M.,
  2014, \mn@doi [\mnras] {10.1093/mnras/stu1391}, \href
  {https://ui.adsabs.harvard.edu/abs/2014MNRAS.444..476R} {444, 476}

\bibitem[\protect\citeauthoryear{{Reid} et~al.,}{{Reid} et~al.}{2016}]{Reid16}
{Reid} B.,  et~al., 2016, \mn@doi [\mnras] {10.1093/mnras/stv2382}, \href
  {https://ui.adsabs.harvard.edu/abs/2016MNRAS.455.1553R} {455, 1553}

\bibitem[\protect\citeauthoryear{{Ross} et~al.,}{{Ross} et~al.}{2020}]{Ross20a}
{Ross} A.~J.,  et~al., 2020, \mn@doi [\mnras] {10.1093/mnras/staa2416}, \href
  {https://ui.adsabs.harvard.edu/abs/2020MNRAS.498.2354R} {498, 2354}

\bibitem[\protect\citeauthoryear{{Rossi} et~al.,}{{Rossi}
  et~al.}{2021}]{Rossi:2020}
{Rossi} G.,  et~al., 2021, \mn@doi [\mnras] {10.1093/mnras/staa3955}, 505, 377

\bibitem[\protect\citeauthoryear{Samushia, Percival  \& Raccanelli}{Samushia
  et~al.}{2012}]{samushia_interpreting_2012}
Samushia L.,  Percival W.~J.,   Raccanelli A.,  2012, \mn@doi [Monthly Notices
  of the Royal Astronomical Society] {10.1111/j.1365-2966.2011.20169.x}, 420,
  2102

\bibitem[\protect\citeauthoryear{Satpathy et~al.,}{Satpathy
  et~al.}{2017}]{satpathy_clustering_2017}
Satpathy S.,  et~al., 2017, \mn@doi [Monthly Notices of the Royal Astronomical
  Society] {10.1093/mnras/stx883}, 469, 1369

\bibitem[\protect\citeauthoryear{{Sheth} \& {Tormen}}{{Sheth} \&
  {Tormen}}{2004}]{Sheth_2004}
{Sheth} R.~K.,  {Tormen} G.,  2004, \mn@doi [\mnras]
  {10.1111/j.1365-2966.2004.07733.x}, \href
  {http://adsabs.harvard.edu/abs/2004MNRAS.350.1385S} {350, 1385}

\bibitem[\protect\citeauthoryear{{Smee} et~al.,}{{Smee}
  et~al.}{2013}]{Smee:2013}
{Smee} S.~A.,  et~al., 2013, \mn@doi [\aj] {10.1088/0004-6256/146/2/32}, \href
  {https://ui.adsabs.harvard.edu/abs/2013AJ....146...32S} {146, 32}

\bibitem[\protect\citeauthoryear{{Smith} et~al.,}{{Smith}
  et~al.}{2020}]{Smith:2020}
{Smith} A.,  et~al., 2020, \mn@doi [\mnras] {10.1093/mnras/staa2825}, \href
  {https://ui.adsabs.harvard.edu/abs/2020MNRAS.499..269S} {499, 269}

\bibitem[\protect\citeauthoryear{Song \& Percival}{Song \&
  Percival}{2009}]{song_reconstructing_2009}
Song Y.-S.,  Percival W.~J.,  2009, \mn@doi [Journal of Cosmology and
  Astroparticle Physics] {10.1088/1475-7516/2009/10/004}, 2009, 004

\bibitem[\protect\citeauthoryear{Sánchez et~al.,}{Sánchez
  et~al.}{2017}]{sanchez_clustering_2017}
Sánchez A.~G.,  et~al., 2017, \mn@doi [Monthly Notices of the Royal
  Astronomical Society] {10.1093/mnras/stw2443}, 464, 1640

\bibitem[\protect\citeauthoryear{{Tamone} et~al.,}{{Tamone}
  et~al.}{2020}]{tamone20a}
{Tamone} A.,  et~al., 2020, \mn@doi [\mnras] {10.1093/mnras/staa3050}, \href
  {https://ui.adsabs.harvard.edu/abs/2020MNRAS.499.5527T} {499, 5527}

\bibitem[\protect\citeauthoryear{{Torrado} \& {Lewis}}{{Torrado} \&
  {Lewis}}{2021}]{torrado:2020xyz}
{Torrado} J.,  {Lewis} A.,  2021, \mn@doi [\jcap]
  {10.1088/1475-7516/2021/05/057}, \href
  {https://ui.adsabs.harvard.edu/abs/2021JCAP...05..057T} {2021, 057}

\bibitem[\protect\citeauthoryear{{Wechsler}, {Zentner}, {Bullock}, {Kravtsov}
  \& {Allgood}}{{Wechsler} et~al.}{2006}]{Wechsler2006}
{Wechsler} R.~H.,  {Zentner} A.~R.,  {Bullock} J.~S.,  {Kravtsov} A.~V.,
  {Allgood} B.,  2006, \mn@doi [\apj] {10.1086/507120}, \href
  {http://adsabs.harvard.edu/abs/2006ApJ...652...71W} {652, 71}

\bibitem[\protect\citeauthoryear{Weinberg, Mortonson, Eisenstein, Hirata, Riess
   \& Rozo}{Weinberg et~al.}{2013}]{weinberg_observational_2013}
Weinberg D.~H.,  Mortonson M.~J.,  Eisenstein D.~J.,  Hirata C.,  Riess A.~G.,
   Rozo E.,  2013, \mn@doi [Physics Reports] {10.1016/j.physrep.2013.05.001},
  530, 87

\bibitem[\protect\citeauthoryear{{Zhai} et~al.,}{{Zhai}
  et~al.}{2017}]{Zhai:2017}
{Zhai} Z.,  et~al., 2017, \mn@doi [\apj] {10.3847/1538-4357/aa8eee}, \href
  {https://ui.adsabs.harvard.edu/abs/2017ApJ...848...76Z} {848, 76}

\bibitem[\protect\citeauthoryear{{Zhai} et~al.,}{{Zhai}
  et~al.}{2019}]{zhai_aemulus3}
{Zhai} Z.,  et~al., 2019, \mn@doi [\apj] {10.3847/1538-4357/ab0d7b}, \href
  {https://ui.adsabs.harvard.edu/abs/2019ApJ...874...95Z} {874, 95}

\bibitem[\protect\citeauthoryear{{Zhai} et~al.,}{{Zhai}
  et~al.}{2022}]{Zhai_2021}
{Zhai} Z.,  et~al., 2022, arXiv e-prints, \href
  {https://ui.adsabs.harvard.edu/abs/2022arXiv220308999Z} {p. arXiv:2203.08999}

\bibitem[\protect\citeauthoryear{{Zhao} et~al.,}{{Zhao} et~al.}{2021}]{zhao20}
{Zhao} C.,  et~al., 2021, \mn@doi [\mnras] {10.1093/mnras/stab510}, \href
  {https://ui.adsabs.harvard.edu/abs/2021MNRAS.503.1149Z} {503, 1149}

\bibitem[\protect\citeauthoryear{{de Mattia} et~al.,}{{de Mattia}
  et~al.}{2021}]{demattia20a}
{de Mattia} A.,  et~al., 2021, \mn@doi [\mnras] {10.1093/mnras/staa3891}, \href
  {https://ui.adsabs.harvard.edu/abs/2021MNRAS.501.5616D} {501, 5616}

\bibitem[\protect\citeauthoryear{{du Mas des Bourboux} et~al.,}{{du Mas des
  Bourboux} et~al.}{2020}]{duMasdesBourboux:2020}
{du Mas des Bourboux} H.,  et~al., 2020, \mn@doi [\apj]
  {10.3847/1538-4357/abb085}, \href
  {https://ui.adsabs.harvard.edu/abs/2020ApJ...901..153D} {901, 153}

\makeatother
\end{thebibliography}




\bsp	
\label{lastpage}
\end{document}